\documentclass[a4paper,fleqn,usenatbib]{mnras}

\usepackage{txfonts}

\usepackage[T1]{fontenc}
\usepackage{ae,aecompl}

\usepackage{graphicx}	% Including figure files
\usepackage{amssymb}	% Extra maths symbols
\usepackage{multirow}
\usepackage{color}
\newcommand{\Msun}{{\rm ~M}_{\odot}}
\newcommand{\Rsun}{{\rm ~R}_{\odot}}
\newcommand{\Zsun}{{\rm ~Z}_{\odot}}
\newcommand{\gpy}{{\rm ~Gpc}^{-3} {\rm ~yr}^{-1}}

\title[Double neutron stars: merger rates revisited]{Double neutron stars: merger rates revisited}

\author[M. Chruslinska et al.]{
Martyna Chruslinska,$^{1,2}$\thanks{E-mail: m.chruslinska@astro.ru.nl}
Krzysztof Belczynski,$^{3}$
Jakub Klencki,$^{1,2}$
\newauthor Matthew Benacquista$^{4}$
\\ 
$^{1}$Astronomical Observatory, University of Warsaw, al. Ujazdowskie 4, 00-478 Warsaw, Poland\\
$^{2}$Institute of Mathematics, Astrophysics and Particle Physics, Radboud University Nijmegen, PO Box 9010, 6500 GL Nijmegen, \\
$^{3}$Nicolaus Copernicus Astronomical Center, Polish Academy of Sciences, Bartycka 18, 00-716 Warsaw, Poland\\
$^{4}$Center for Gravitational Wave Astronomy, University of Texas Rio Grande Valley, Brownsville, TX, USA
}
\date{Accepted XXX. Received YYY; in original form ZZZ}

\pubyear{2017}
\begin{document}
\label{firstpage}
\pagerange{\pageref{firstpage}--\pageref{lastpage}}
\maketitle
\begin{abstract}
We revisit double neutron star (DNS) formation in the classical binary evolution 
scenario in light of the recent LIGO/Virgo DNS detection (GW170817). 
The observationally estimated Galactic DNS merger rate of 
$R_{\rm MW}=21^{+28}_{-14}$ Myr$^{-1}$, based on 3 Galactic DNS systems,
fully supports our standard input physics model with $R_{\rm MW} =24$ Myr$^{-1}$. 
This estimate for the Galaxy translates in a non-trivial way (due to cosmological 
evolution of progenitor stars in chemically evolving Universe) into a local 
($z\approx0$) DNS merger rate density of $R_{\rm local}=48$ Gpc$^{-3}$yr$^{-1}$, 
which {\em is not} consistent with the current LIGO/Virgo DNS merger rate
estimate ($1540^{+3200}_{-1220}$ Gpc$^{-3}$yr$^{-1}$).
Within our study of the parameter space we find solutions that allow for 
DNS merger rates as high as $R_{\rm local} \approx 600^{+600}_{-300}$ Gpc$^{-3}$yr$^{-1}$ 
which are thus consistent with the LIGO/Virgo estimate. 
However, our corresponding BH-BH merger rates for the models with high DNS merger rates 
exceed the current LIGO/Virgo estimate of local BH-BH merger rate ($12$--$213 \gpy$). 
Apart from being particularly sensitive to the common envelope treatment, DNS 
merger rates are rather robust against variations of several of the key factors 
probed in our study (e.g. mass transfer, angular momentum loss, natal kicks). 
This might suggest that either common envelope development/survival 
works differently for DNS ($\sim 10-20 \Msun$ stars) than for BH-BH ($\sim 40-100 \Msun$ stars) progenitors, or high BH natal kicks are needed to meet observational constraints for both
types of binaries. 
Note that our conclusion is based on a limited number of (21) evolutionary models and is valid only 
within this particular DNS and BH-BH isolated binary formation scenario.
\end{abstract}


%%%%%%%%%%%%%%%%%%%%%%%%%%%%%%%%%%%%%%%%%%%%%%%%%%

%%%%%%%%%%%%%%%%% BODY OF PAPER %%%%%%%%%%%%%%%%%%

\section{Introduction}

The first ever detection of the gravitational wave signal from coalescing double compact objects \citep{LIGO16} by
the Laser Interferometer Gravitational-wave Observatory (LIGO) opened a completely new window on the Universe,
providing scientists with a new tool to study the origin of these objects.
With increasing sensitivity and a growing number of operating detectors this tool will bring more information leading  
us closer to the correct understanding of the physics standing behind their evolution and formation.
Each new detection or further non-detection will impose stronger observational limits on the merger rates.
This in turn will help to discriminate between evolutionary paths that predict too many or too few detections and better constrain the uncertain
phases of the evolution of compact binaries. At the same time, theoretical predictions of the properties of 
 double compact objects merging within the Hubble time (the merging population hereafter) can serve to better tune the detectors,
 making the search more efficient.
It is thus important to identify the factors that affect the rates and the properties of the merging population the most.
\\
In this study we focus on merging double neutron star (DNS) binaries.
There are two essentially different formation channels for merging DNS: isolated 
\citep[field populations, e.g.][]{ZwartYungelson98,BelczynskiKalogeraBulik02,VossTauris03,Belczynski07,Bogomazov2006,Dominik12,Dominik13,Dominik15,MennekensVanbeveren14,TaurisLangerPodsiadlowski15,
BelczynskiRepetto16,Belczynski16N}
where two stars are born together and their evolution is affected only by the interaction between the binary components, 
and dynamical - occurring in dense environments, where the dynamical interactions between stars are important e.g. in globular clusters or nuclear clusters
(e.g. \citealt{Phinney96}; \citealt*{Grindlay06}; \citealt{Ivanova08}; \citealt{Petrovich17}).
We consider only the isolated channel (from now on we refer only to the DNS that formed in isolation, unless explicitly stated otherwise),
aiming to pinpoint the crucial factors that lead to the biggest uncertainties of theoretically predicted rates.
\\
A theoretical, population synthesis-based approach, suffers from a number of uncertainties, arising from different sources: the
initial conditions used in the simulations, the input physics, the treatment of weakly constrained phases of stellar and binary evolution and assumptions
about e.g. the history of the star formation and metallicity on the cosmological timescales, necessary to normalize the merger rates.
\citet{deMinkBelczynski15} examined the impact of the updated initial binary parameter distributions obtained by 
\citet{Sana2012} from the spectroscopic measurements of massive O-type stars in young (a few Myr) stellar clusters and associations on the merger
rate predictions. 
They conclude that the updated distributions from which the initial orbital parameters are chosen and 
variations applied to parameters of the distributions within the observational limits have only a marginal effect
on the coalescence rates, unless the slope of the high-mass initial mass function (IMF) is considered, which can shift the rates by a factor of a few.
Recently, \citet{MoediStefano16} motivated by the recent observational studies analyzed the data from numerous surveys of massive binary stars
and revised the initial distributions, finding statistically significant correlations between many of the physical parameters of the binaries
to date treated separately.
The study of the effect of these inter-correlated initial binary parameters on the coalescence rates is underway (Klencki et al. in prep.).
\\
Here we concentrate on the uncertainties related to the stellar evolution and binary interaction. 
The description of all of the crucial phases of the evolution of the progenitor of a merging double neutron star system requires making certain assumptions
often expressed in the form of a particular parametrization and deciding upon a particular value of the free parameters, not always well
constrained by observations  \citep[with an infamous example of the common envelope, e.g.][]{Ivanova13}.
The impact of those choices on the final result (e.g. the estimated merger rate) was studied to some extent by a number of authors
and with the use of different population synthesis codes (e.g. \citet{ZwartYungelson98, VossTauris03,Bogomazov2006,MennekensVanbeveren14})
also with the \textsc{StarTrack} code \citep{Belczynski2002, Belczynski2008}, e.g. the study of \citet{Dominik12} focusing on the common envelope phase),
which we use in this work.
However, here we aim to make the whole picture a little bit more complete, examining the impact on the merger rates of the factors
related to the recent observational and theoretical progress in the study of the evolution of double compact objects,
paying particular attention to the NS-NS binaries.\\

\section{Evolution towards the DNS: crucial phases and bottlenecks}\label{sec: assumptions}

The typical isolated evolution of the progenitor of a merging DNS binary requires the occurrence of two non-disrupting core-collapse events,
the interaction between the binary components via mass transfer phase(s) and the presence of the mechanism capable of severely decreasing the orbital separation
that will not lead to the coalescence before the formation of two compact objects.
In this section we discuss the uncertainties related to those phases and the results of the recent studies.\\

\subsection{Core-collapse events and the natal kicks}
The progenitor of a merging DNS system encounters two core-collapse events and the first set of 'evolutionary' uncertainties is related to the assumptions 
made to describe the mechanism of the explosion/collapse, e.g. criteria used to identify the progenitors of stars undergoing different types of core-collapse
(i.e. iron-core collapse supernova, electron-capture supernova \citep[e.g.][]{Miyaji80}, accretion induced collapse \citep{Nomoto91}) 
and the magnitude of the related natal kick velocity imparted on the newly-formed neutron star (NS) \citep[e.g.][]{Gunn70,Hobbs2005}.
Natal kicks are believed to arise from asymmetries developed during the explosion/collapse, expressed in either asymmetrically ejected mass 
\citep{Janka_Muller94,Wongwathanarat13,Janka17} or anisotropic emission of neutrinos \citep[e.g.][]{Fryer_Kusenko06} (or due to both factors acting together).
\\
With an intention to statistically describe the natal kicks, \citet{Hobbs2005} studied the proper motions of young (with characteristic ages of less than 3 Myr)
and presumably single field pulsars. They found that the distribution of their velocities can be well described by a Maxwellian distribution with the velocity
dispersion of $\sigma$ = 265 km/s, which yields an average natal kick magnitude of 420 km/s.
At the same time, there is mounting evidence that some NS form with substantially smaller natal kicks. For instance, \citet{Pfahl02} find that the orbital
parameters of a certain class of the high-mass X-ray binaries distinguished by relatively long orbital periods (in excess of 30 days) and small eccentricities
(e$\lesssim$ 0.2) suggest that neutron stars in these systems must have formed with birth velocities smaller than $\sim$50 km/s.
Also the orbital parameters of some of the Galactic double neutron stars seem to require low natal kick at birth \citep{Heuvel07,BeniaminiPiran16}.
Another commonly quoted argument is the retention of NS in globular clusters \citep{Pfahl02gc,Ivanova08}.
As these structures typically have escape velocities around 50 km/s, larger kicks would make it impossible to retain neutron stars inside their potential well.
However, there are many pulsars observed within the globular clusters and we know of two double neutron star binaries (B2127+11, J1807-2500) contained within them.
\\
To explain the origin of these low-kick neutron stars it was postulated that they form when a degenerate ONeMg collapses due to electron-capture reactions,
as opposed to a standard iron-core core-collapse supernova (CCSN), either in an electron-capture supernova (ECS) or accretion-induced collapse (AIC)
of an accreting, massive white dwarf (WD) to a NS.
\\
Different structures of the progenitors of ECS(AIC) and CCSN just before the explosion
are believed to lead to differences in the dynamics of the supernova (collapse) itself.
Progenitors of ECS/AIC are claimed to collapse more rapidly than iron-core collapsing stars
due to much steeper density gradients at the edge of the core, leading to lower explosion energies
and smaller masses of the ejected material \citep{Dessart06,Jones13,Schwab15}.
The most important consequence of these differences is expressed in the magnitude of the natal kick gained by a newly formed neutron star.
In the rapid explosion substantial asymmetries may not be able to develop. 
Natal kicks are commonly assigned to arise due to these asymmetries, hence the smaller birth velocities expected for ECS than normal CCSN. 
However, if this differentiation is truly determined by the density gradient at the edge of the collapsing core,
lowest mass CCSN progenitors may also form NS with small natal kicks, as their density profiles resemble those
expected for ECS progenitors \citep{Jones13}.\\
It has been also suggested that the explosions of heavily stripped massive stars which have lost their envelopes via binary interactions 
 may lead to small natal kicks, especially if kicks are primarily due to asymmetries in ejected mass
 and gravitational interaction of the ejecta with a just-formed NS rather than anisotropic neutrino emission, as argued by \citep{Wongwathanarat13}.
 The extreme stripping of the envelope may arise due to interaction with a compact companion star,
 as in the scenario described by \citet{Tauris13,TaurisLangerPodsiadlowski15,Moriya17} (leading to so-called ultra-stripped supernovae; see \ref{ap:USSN}), 
 or with a non-compact companion via the double-core scenario \citep{Brown95} as in the variation described by \citet{Belczynski01}, involving the evolution
 of stars of similar initial masses that pass through the common envelope combined from the envelopes of both companions and leaving behind the bare CO cores
 in a compact orbit evolving in parallel.
 All the ultra-stripped SN scenario, the double core scenario and electron-capture collapse mechanisms predict that low-kick neutron stars would form mainly (if not only) in binaries
 \citep{Podsiadlowski04,Heuvel07}.
  \\ \newline
 The standard treatment of natal kicks in population synthesis studies relies on single or multi-component Maxwellian distributions to describe the magnitudes of the
 velocities. In \textsc{StarTrack} we use the form proposed by \citet{Hobbs2005} as a default. However, in such an approach there is no link between the properties
 of an exploding star and the velocity that is acquired by the remnant. A number of authors suggested that there should be a connection between the magnitude of the kick
 and the mass of the ejected material, mass of the remnant NS and/or mass of the collapsing core \citep{Jones13,TaurisLangerPodsiadlowski15,BeniaminiPiran16,BrayEldridge16,Janka17}.
 Recently \citet{BeniaminiPiran16} (based on the analysis of the observed properties of known DNS binaries) and \citet{BrayEldridge16}
 suggested a simple, linear relationship between the natal kick velocity and the ratio of the mass ejected during the SN to NS mass, which was further
 discussed by \citet{Janka17}.
 Nonetheless, the form of this connection is not clear, neither is the role of neutrinos in the natal kick development.
 If it proves to be significant, the above considerations may turn out erroneous.
 
 \subsection{Progenitors of the electron-capture supernovae}
 
  The magnitudes of the natal kicks as described by the distribution proposed by Hobbs et al.  
 pose a challenge for the formation of double neutron stars (and binaries with neutron stars in general), as such velocities are high enough to disrupt the binary,
 especially at the first supernova when the first compact remnant is formed, as then the orbit is still relatively wide.
 One can thus presume that the abundance of the merging population of DNS would be sensitive to the assumptions regarding the 
 electron capture supernovae and AIC channels, potentially allowing for the formation with a small birth velocity.
 \\
 \newline
 The ECS/AIC progenitors originate from stars with initial masses from the 'transition region' (around 8-9 $\Msun$ at solar metallicity) between stars that finish their evolution as
 a WD and NS and form degenerate ONeMg cores after the carbon burning phase.
When such a core approaches the Chandrasekhar mass limit (1.375 $\Msun$) electron captures on Mg and Ne can trigger the collapse to a NS 
\citep[e.g.][]{Miyaji80, Nomoto91,Jones16}.
\\ \newline
For single stars the range of the initial masses allowed for ECS progenitors is very narrow and nearly all NS forming in ECS are believed to originate in binary systems,
where the interaction (via mass transfer phases) with a companion star can broaden this range, so that initially less or more massive
stars can end up with the same core structure before the supernova explosion \citep{Podsiadlowski04,Heuvel07}.
However, it is not clear how wide this range can be, in particular the location of the lower limit on the mass of a He core at the beginning of the AGB stage
for a potential ECS progenitor (or lower and upper limit for the zero-age main sequence mass of an ECS progenitor).
The upper limit seems to be around 2.5$\Msun$ and is due to the fact that in the heavier cores the oxygen will always be ignited and the star will proceed
with the burning of the heavier elements, until an iron core is formed.
In \textsc{StarTrack} we follow the study of \citet*{Hurley2000} \citep[with the updated mass range based on the results of][see \ref{sec: method:ECS} for details]{EldridgeTout04a,EldridgeTout04b} and
use the mass of the core at this stage to decide whether a star will develop a degenerate ONeMg core and thus whether at some point it may collapse due to
electron-capture processes (either ECS or AIC).
The determination of those limits requires detailed stellar structure modeling and many factors may affect the range, even without invoking the binary evolution.
The most important factors seem to be
the amount of convective core overshooting and rotational mixing,
the treatment of convection and mixing in general, the efficiency of the second dredge-up, the mass loss rate, even the nuclear reaction rates
\citep{Poelarends08,Jones16,Doherty17}. 
In case of binary stars these are also the initial pericentre distance, which
for the fixed mass ratio has a decisive role in determining the type/moment of the interaction (mass transfer) between the components,
as well as assumptions about the conservativeness of the mass transfer.

\subsection{Mass loss and angular momentum loss}

The merging population of double neutron stars, out of necessity, originates from binaries that are at some point of their evolution close enough
to encounter the mass transfer (MT) phase/phases (unless the wide, non-interacting binary was severely narrowed due to the favorable direction of the natal kick
during the second supernova). These phases, as well as the stellar winds which are important for the evolution of O/B type progenitor stars of NS, require
specifying how much mass is lost from the system and how much angular momentum is carried away by the escaping material, which is crucial for the
evolution of the orbital separation during the MT.
The latter depends on the adopted physical model of how the matter escapes from the binary - whether it is from the vicinity of the accretor or donor,
with relatively low angular momentum loss or it occurs through the second Lagrangian point $\rm L_{2}$ or is liberated 
from a circumbinary disc of a certain radius, leading to high angular momentum loss.
In theoretical studies the angular momentum loss during the MT is usually parametrized by the  quantity $j_{loss}$ standing for the fraction of
 angular momentum lost from the system with the escaping matter. 
 The second commonly used mass transfer parameter describes the fraction of the mass lost by the donor star that is accreted by its companion
 (conservativeness $f_{a}$, where $f_{a}$=1 means fully conservative MT).
 As argumented by \citet*{deMink07}, using a single value of $f_{a}$ in a population synthesis study, as we assume in \textsc{StarTrack} using $f_{a}$=0.5 as a default, is a very
 crude approximation.
 However, the degree to which MT is conservative is rather poorly constrained observationally \citep{deMink07,Pols07}.
\\
The mass lost by stellar winds in population synthesis studies is usually assumed to be spherically symmetrical and homogeneous \citep[e.g.][]{Belczynski2008,MennekensVanbeveren14},
relying on the mass loss rate description by \citet*{Vink01} or \citet{deJager88} for O/B type stars and carrying the specific angular momentum of the mass losing star.
However, arguments were raised that the wind may in fact be permeated by small-scale inhomogeneities (so-called clumps) and the above mentioned prescriptions, which are based on
the assumption of homogeneous wind lead to main sequence mass-loss rates that are overestimated by a factor of 2-3 \citep[see sec. 2 in][and references therein]{Smith14}.
At the same time the wind mass-loss rates for the red supergiants may be higher than it follows from the \citet{Vink01} prescription for hot stars \citep{vanLoon00},
affecting the late evolution of $\sim$8-35 $\Msun$ NS progenitor stars.

\subsection{Common envelope with Hertzsprung-gap donors}\label{sec: assumptions:CE}

The common envelope (CE) phase is probably the most problematic stage of the binary evolution, as the criteria for the formation and ejection of the 
envelope are very uncertain \citep[see e.g.][for the review]{TaamSandquist00,Ivanova13}.
At the same time, it provides the mechanism to shrink the orbital separation by even a few orders of magnitude,
the decrease that is needed to explain the formation of some of the observed close systems \citep*[e.g. low mass X-ray binaries][]{Liu07}
and account for the formation of merging double compact objects without dynamical interactions \citep*[e.g.][]{Mapelli09,Rodriguez16} or rapid stellar 
rotation \citep{deMinkMandel16,Marchant16}. 
The radii of progenitors of compact stars in these binaries typically reach 100 - 1000 $\Rsun$, which is much larger than the required separation
for the merging systems ($\lesssim$ a few solar radii for DNS binary).\\
In general, CE forms when the mass transfer rate is too high for the accretor to accommodate all of the 
transferred material (MT is unstable), giving rise to a short-lived phase during which both stars orbit inside a shared envelope.
This is believed to lead to a binary spiral-in due to increased friction and, if the envelope is not ejected beforehand (e.g. at the expense of the orbital energy),
may lead to its coalescence that aborts further binary evolution and potential DNS formation.\\
CE initiated by donor stars at the Hertzsprung-gap (HG) deserves a particular attention. 
It is believed that due to lack of the clear distinction between the core and the envelope structure \citep[no steep density drop at the boundary, see][]{IvanovaTaam04},
the common envelope initiated by HG stars most likely leads to premature merger and hence the
significant reduction in the expected double compact object merger rates \citep{Belczynski07}.
We refer to stars being on the HG in the same manner as Hurley et al. (2000).
Thus, we consider both the stars with hydrogen-rich envelopes transitioning to the red giant branch after the core hydrogen exhaustion and helium-rich stars at the end
of the core helium burning phase that have lost their hydrogen envelopes (e.g. due to binary interaction or strong stellar winds; so-called Helium-Hertzsprung Gap).
As a substantial fraction of the merging binaries passes through this kind of CE phase, allowing for its survival gives rise to the optimistic (the highest)
estimate for the rates \citep[e.g.][]{Dominik12,Dominik13,Dominik15,deMinkBelczynski15,Belczynski16N}.
However, these optimistic rates are not supported by the current LIGO estimates of BH-BH merger rates \citet{Abbott17_GW170104}.
The recent investigations by \citet{PavlovskiiIvanova15} and \citet{Pavlovskii17} revealed that in some cases where based on earlier studies one would expect 
the common envelope initiated by a massive convective giant donor to ensue, the mass transfer may in fact be stable.
This might potentially lead to a much higher orbital separation after the mass transfer.
In \citet{Pavlovskii17} they focus on the mass transfer in progenitor binaries of close BH-BH systems with massive HG donors and stellar-mass BH
and conclude that even despite the extreme mass ratio the system evolves through stable MT instead of CE and does not form a merging binary.
However, so far their study, and thus the validity of the conclusions, are limited to only a few values of metallicities and cases of MT with compact
accretor (i.e. companion star that is far from filling its Roche lobe).

\section{Computational method}

We follow the evolution of a binary starting with both stars at zero-age main sequence (ZAMS) until the eventual formation of two compact objects,
using the \textsc{StarTrack} code. The code is based on the fit formulas to the detailed evolutionary models developed by \citet{Hurley2000} and described in detail
in \citet{Belczynski2002,Belczynski2008}. Since then the code has undergone a few major upgrades.
The current treatment of the stellar wind mass-loss is described in section 2 in \citet{Belczynski2010a}.
For the common envelope evolution we use the energy-based formalism of \citet{Webbink84} with the parameters $\alpha$ 
(efficiency of the conversion of orbital energy into unbinding the envelope, always set to 1) and $\lambda$ (the binding energy of the envelope) dependent
on the ZAMS mass of the donor star, its evolutionary stage, radius and the mass of the envelope, using the estimates made by
\citep[][see \citet{Dominik12} for the details]{XuLi10}.
The parameters of a binary at ZAMS (mass of the primary, mass ratio, eccentricity, period) are chosen from the distributions described in \citet{Sana2012},
based on spectroscopic observations of Galactic O-type stars.
These include the eccentricity distribution $f(e) \propto e^{-0.42}$ in the range $[0.0 \, , \, 1.0]$,
period distribution $f(P) \propto \rm log(P)^{-0.55}$ in the range $[0.15 \, , \, 5.5]$, 
a Kroupa-like IMF\footnote{ We refer to the IMF function defined as the number of stars with a birth mass 
within a certain mass range using linear mass bins $dN/dm \sim m^{-\alpha}$, 
in contrast to the logarithmic bins $dN/d log(m) \sim m^{-\Gamma}$ used by some authors. 
Note that $\alpha = \Gamma + 1$} \citep*[][]{Kroupa1993}:
\begin{equation} \label{imf}
dN/dM \propto \left\{
\begin{array}{l c}
M_a^{-1.3} & \quad 0.08 \ \Msun \leq M_1 < 0.5 \ \Msun \\
M_a^{-2.2} & \quad 0.5 \ \Msun \leq M_1 < 1.0 \ \Msun \\
M_a^{-\alpha_{\rm IMF}} & \quad 1.0 \ \Msun \leq M_1 < 150 \ \Msun, \\
\end{array}
\right.
\end{equation}
with the high-mass ($\rm M> 1.0 \, M_{\odot}$) power-law exponent of $\alpha_{\rm IMF}=2.3$, from which the mass of the more massive star from
the binary is drawn and a flat binary mass ratio $q = \rm M_{\rm b} / \rm M_{\rm a}$ distribution in the range $[0.0 \, , \, 1.0]$ from which we choose
the mass of the less massive component.
Note that the high-mass IMF exponent is different than in \citet{Dominik12,Dominik13,Dominik15,deMinkBelczynski15,BelczynskiRepetto16} and earlier studies employing 
\textsc{StarTrack}, where the high-mass IMF slope of 2.7 was used. 
This change was motivated by the observational results of the IMF studies \citep[e.g.][]{Bastian10,Offner14,Krumholz14}.
As shown by \citet{deMinkBelczynski15}, this leads to higher rate estimates (by a factor of $\sim$3-6 for NS-NS and BH-BH binaries respectively when $\alpha_{\rm IMF}$
was changed from 2.7 to 2.2 ) than in those earlier studies.

\subsection{The reference model and applied variations} \label{sec: ref_model}

Apart from the different initial distributions and $\alpha_{\rm IMF}$ our reference model is identical to the standard model in \citet{Dominik12}
and follows the assumptions discussed in the papers describing the code \citet{Belczynski2002,Belczynski2008}. Below we shortly introduce the relevant
standard assumptions and variations applied to this standard approach, motivated by the results of studies discussed in section \ref{sec: assumptions}.
In other models only the factors explicitly given are varied, the rest is kept the same as in the reference model.
We simulated $2 \cdot 10^{7}$ ZAMS binaries for each model and each metallicity, unless the other number is provided in the model description.
We draw the mass of the primary from the range $4<M_{a}<150 \Msun$ and from $0.08<M_{b}<150 \Msun$ for the secondary.
Our models are summarized in table \ref{tab:models}.

\subsubsection{Core-collapse events and the natal kicks}

The treatment of the SN explosion is based on a study by \citet{Fryer12} (the 'rapid' supernova model that reproduces the observed mass gap between NS and BH,
\citep{Belczynski12}).
For CCSN-formed NS the magnitudes of the natal kicks are drawn from the Maxwellian distribution 
with the velocity dispersion $\sigma$=265 km/s, as proposed by \citet{Hobbs2005}, with a random direction.
For compact objects formed with partial fallback the magnitude of the kick is lowered according to the formula: $V_{kick} = (1-f)V_{nk}$, where $V_{nk}$ is the 
 velocity drawn from the distribution and $f$ is the fraction of the mass of the stellar envelope that fall back.
Fallback is important only for BH and the most massive NS.
The most massive BH form via direct-collapse, with zero natal kick \citep[see][for the detailed discussion of BH kicks in \textsc{StarTrack}]{BelczynskiRepetto16}.
We allow for the formation of the NS via ECS and AIC. In our standard approach no natal kick is added to such a NS.
However, if there is a mass ejected from the system, the centre of mass of a binary still gains an additional velocity after the SN \citep[the Blaauw kick;][]{Blaauw61}
\\ \newline
%We calculate a set of models where the assumptions about the natal kicks are varied.
In the $BE$ set of models, instead of using a Maxwellian distribution to describe the magnitude of natal velocity, which puts aside any
potential connection between the kick and the properties of an exploding star, we adopt the prescription proposed by \citet{BrayEldridge16} (models called $BE$ after the names of the authors).
In their study NS birth velocities are assumed to be a result of the asymmetric ejection of the envelope and
the conservation of momentum between this and the newly formed neutron star:
\begin{equation}\label{eq: BE}
 v_{nk}= \alpha \frac{M_{ej}}{M_{rem}}+ \beta
\end{equation}
$\alpha$ and $\beta$ are assumed to be universal for all supernovae (however, ECS may require a separate treatment).
The first constant represents the net velocity gained by the remnant star due to asymmetric ejecta, while the second term
is interpreted as representing any potential contribution to the natal kick (at the same direction) from another source (e.g. anisotropic neutrino emission).
We use the coefficients corresponding to their best fit to the Hobbs 3D distribution: $\alpha=70 \rm \ km/s$ and $\beta=120 \rm \ km/s$ to draw
the natal kick velocities after CCSN ($BE1$ model), after both CCSN and ECS ($BE2$) and after CCSN, ECS and AIC NS formation ($BE3$).
We also test the relation (\ref{eq: BE}) with coefficients given by their lower limits: $\alpha=60 \rm \ km/s$ and $\beta=70 \rm \ km/s$ ($BE4$,$BE5$).
Note that for ECS formed NS this is probably an exaggerated kick, as they are believed to form with velocities $\lesssim 50 \ \rm km/s$ \citep[e.g.][]{Pfahl02},
even less certain is the eventual kick for AIC-formed NS.
In model $NK1$ we keep the standard assumptions about the CCSN natal kick, but add a small birth velocity after ECS, drawn from Hobbs-like distribution
with $\sigma$=26.5 km/s.
In model $NK2$ we depart from the standard Maxwellian distribution to allow more CCSN to lead to small natal kicks.
We combine the flat distribution extending to velocities $v_{nk}\leqslant$50 km/s with Hobbs et al. distribution ($\sigma \rm = 265 \ km/s$) for $v_{nk}>$50 km/s.
The distributions are combined in such a way that it is equally probable that the natal kick magnitude is drawn from the flat and the Maxwellian part.

\subsection{Progenitors of the electron-capture supernovae} \label{sec: method:ECS}
To decide whether a star explodes as an electron-capture SN we use its He core mass ($\rm M_{He}$) at the beginning of the AGB stage
to constrain the fate of its CO core, following the study of \citet{Hurley2000}, using masses calculated with the same,
but updated evolutionary code \citep{EldridgeTout04a,EldridgeTout04b}.\\
If $\rm M_{He} <$ M$_{\rm cbur1} = 1.83 \Msun$ the star forms a degenerate CO core and finishes its evolution as a CO WD. 
If $\rm M_{He} >$ M$_{\rm cbur2} = 2.25 \Msun$ a non-degenerate CO core is formed and the star proceeds with burning 
of heavier elements until an iron core forms and collapses to a NS/BH.
Stars with M$_{\rm cbur1}$ $> \rm M_{He} >$ M$_{\rm cbur2}$ form partially degenerate CO cores.
If such a core reaches the critical mass of 1.08 $\Msun$ (Hurley et al. 2000), it ignites CO off-centre and non-explosively burns CO into ONe,
forming a degenerate ONeMg core. If this core reaches M$_{\rm ecs} = 1.38 \Msun$ it collapses to a NS in an ECS, otherwise
an ONeMg WD is formed, which still can collapse to a NS if its core exceeds M$_{\rm ecs}$ due to accretion of mass
during a subsequent RLOF phase (AIC channel). \\
The NS formed in ECS or AIC are assigned the masses of M$_{\rm rem,bar}$ = M$_{\rm ecs}$, corresponding to gravitational mass of 1.26 $\Msun$.
In model $EC$ we widen the mass range for the He core that can evolve into a degenerate ONeMg core, setting M$_{\rm cbur1} = 1.63$ and M$_{\rm cbur2} = 2.45$.
\subsection{Mass loss and angular momentum loss}
The standard assumptions about the stable mass transfer are outlined in section 3.4 in \citet{Belczynski2008}.
For non-compact accretors the angular momentum loss during the MT is governed by the equation:
\begin{equation}
 dJ_{\rm RLOF}/dt = j_{loss} \frac{J_{\rm orb}}{M_{\rm don}+M_{\rm acc}}(1-f_{a})\dot{M}_{\rm don}
\end{equation}
where the dimensionless coefficient $j_{loss}$ describes the fraction of angular momentum carried away by the ejected material,
$f_{a}$ describes the conservativeness of the MT and
$M_{\rm don}$ and $M_{\rm acc}$ are the masses of the donor and accretor respectively. In the standard model $j_{loss}$=1 and $f_{a}$=0.5 (half-conservative MT)
in all cases.\\
In model $J1$ we assume extreme angular momentum losses $j_{loss}$=5, while in $J2$ we use lower $j_{loss}$=0.2. 
We also introduce variations $J3$ and $J4$, which are non-conservative
($f_{a}$=0) versions of models $J1$ and $J2$ respectively. The conservative case (model $J5$) $f_{a}$=1 is also shown (although it was previously considered in \citet{Dominik12}).
In the case of stable MT with compact accretors (WD, NS, BH) the accretion is limited to the Eddington critical rate \citep[eq. 30 in][]{Belczynski2008}
If the mass transfer rate is higher than this critical rate, the excess material is assumed to escape the system from the vicinity of the accretor.
The associated angular momentum loss is then given by:
\begin{equation}\label{eq:compact_Jloss}
  dJ_{\rm RLOF}/dt = \frac{2\pi}{P_{\rm orb}}R^{2}_{\rm com}(1-f_{a})\dot{M}_{\rm don}
\end{equation}
where $P_{\rm orb}$ is the orbital period and $R_{\rm com}=aM_{\rm don}/(M_{\rm don}+M_{\rm acc})$ is the distance between the accretor and centre of mass.
If the Eddington rate $\dot{M}_{\rm edd}$ is not exceeded then MT it is conservative, otherwise $f_{a}$ = $\dot{M}_{\rm edd}/\dot{M}_{\rm don}$.
In model $J6$ we multiply eq. \ref{eq:compact_Jloss} by 0.2, lowering the angular momentum losses from MT with accreting NS, 
while in $J7$ we increase them by a factor of 2.\\
We also introduce two models with varied wind mass-loss rates. In $W1$ we lower the rates by a factor of 2 for main sequence stars, while increasing
them by a factor of 2 for helium-rich stars (mass loss rates for luminous blue variables are kept the same as in the standard model).
\citet{Dominik12} presented the model with lowered wind mass loss rates in both cases. 
Although, as discussed in section \ref{sec: assumptions}, the recent findings favor lower main sequence wind mass-losses,
we include also the variation $W2$ with the rates increased
by a factor of 2 for completeness.

\subsection{Common envelope with Hertzsprung-gap donors}
As the fate of the binaries undergoing common envelope initiated by Hertzsprung-gap stars is not yet clear, 
for each model and metallicity we show the merger rate computed with earlier exclusion of such binaries (submodel $B$) and
the case where we allow them to evolve through CE with our standard assumptions (submodel $A$).\\
Additionally, in model $P$ we force the binary which is about to start the CE evolution 
\citep[according to our standard criteria, see sec. 5 in][]{Belczynski2008}
with HG donor and a NS or BH accretor to initiate a stable thermal mass transfer instead.
\newline
\\
We also introduce two 'combo' models: $C$ which combines the factors that are expected to lead to increased merger rates for double neutron stars, 
that is the wider ECS channel $EC$, $BE1$ model for the natal kicks (CCSN-born NS natal kick $\propto M_{ej}/M_{rem}$) and low angular momentum losses $J2$ 
and $C+P$ model which apart from all of the variations combined in $C$ includes also $P$ variation.

\subsection{The merger rates} \label{sec: rates}
In this work we use three different quantities to provide the estimate of the merger rates.
The first requires no additional assumption and is simply the number of double neutron stars merging within the Hubble time in a given model and 
for a given metallicity.
Secondly, we present Galactic merger rates (number of coalescences per unit time within a galaxy) calculated for a fiducial Milky Way-like galaxy 
(assuming 10 Gyr of continuous star formation at the rate of 3.5 $\Msun$/yr) for a single metallicity, using the same approach as described in 
section 4 in \citet{Dominik12}.\\
For each model we present the results for three metallicities (solar: $Z_{\odot}$=0.02 \citep{Villante14}, 10\% $Z_{\odot}$ and 1\% $Z_{\odot}$),
to show the variation of the merger rates with this property.\\
As a third estimate we provide the local merger rate density (number of coalescences per unit time per unit volume in the local Universe, i.e. around redshift=0),
calculated with taking into account the star formation rate and metallicity evolution with redshift.
We adopt the cosmic star formation rate SFR(z) as in \citet{Madau14} and metallicity evolution with redshift following the estimate of \citet{Madau14},
but increased by 0.5 dex to better fit observational data \citep[][see eq. (2) in Belczynski et al. 2016a ]{Vangioni15}.
See appendix \ref{ap: rates} for the details.
\\
The local rates can be approximately translated to the expected LIGO detection rates using the formula (see sec. III in \citet{Abadie2010} and sec. 3 in \citet{Dominik15}):
\begin{equation}\label{eq: detection}
 R_{det} = \frac{4 \pi}{3} D_{h}^{3} w^{3} \left< (\mathcal{M}/1.2 \Msun)^{15/6} \right> R_{\rm local}
\end{equation}
where $w\approx (2.264)^{-1}$, $D_{h}$ is the horizon distance (the luminosity distance at which an optimally oriented canonical (1.4 +1.4)$\Msun$ NS-NS binary would be detected at 
a signal-to-noise ratio SNR=8), $R_{\rm local}$ is the local merger rate density
and $\mathcal{M}=(M_{a}+M_{b})\cdot (M_{a} M_{b}/(M_{a}+M_{b})^{2})^{3/5}$ is the chirp mass, where $M_{a}$ and $M_{b}$
are the final masses of the binary components of the detected merging system.
The typical value of $\mathcal{M}$ for the merging population of DNS binaries is around 1.06 - 1.15 for different models.
The expected advanced LIGO (assuming the full design sensitivity) $D_{h}$ during the O3 run for DNS is $\sim$271-385 Mpc \citep{Abbott16b}.
We calculate the approximate DNS detection rates using equation \ref{eq: detection}, calculated local merger rate density and 
the maximum O3 $D_{h}$=385 Mpc and advanced LIGO $D_{h}$=487 Mpc for DNS \citep[corresponding to DNS range 170 Mpc and 215 Mpc respectively]{Abbott16b}.
The selected models are available on the Synthetic Universe website \footnote{http://www.syntheticuniverse.org/}.

\section{Results}

\begin{figure}
	\includegraphics[width=\columnwidth]{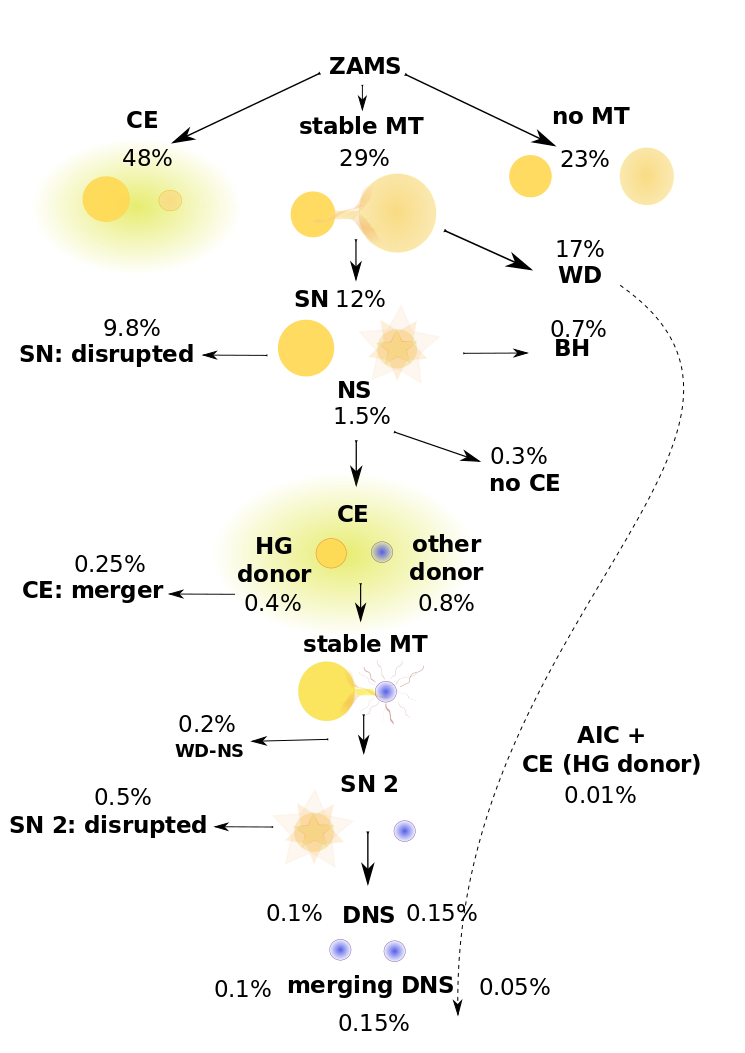}
    \caption{
      The leading evolutionary channel (in the middle) for the formation of the merging double neutron star system, shown for the reference model
      (\ref{sec: ref_model}) and solar metallicity ($\Zsun$=0.02),
      simulated with the ZAMS masses from the range $4 \Msun <M_{a}<150 \Msun$ and $0.08 \Msun <M_{b}<150 \Msun$.
      The other channels (with the common envelope (CE) or no mass transfer (MT) at all before the first supernova (SN))
      produce at most an order of magnitude less merging double neutron stars (DNS).
      We differentiate between the two CE cases: with a Hertzsprung gap (HG) donor (percentage on the left)
      and with any other donor (percentage on the right). The latter case constitutes our submodel $B$, while both contribute to submodel $A$.
      All numbers refer to the fraction of the simulated ZAMS binaries. See text and appendix \ref{ap: evolution} for details.
    }
    \label{fig:ref02}
\end{figure}

Only a small fraction of the simulated ZAMS binaries ends up forming a double neutron star system that will merge within the Hubble time.
Many of them get disrupted during the SN, merge during the CE phases or form other types of binaries.
The fate of the simulated systems is summarized in the figure \ref{fig:ref02}
which also schematically shows the crucial phases of the evolution of the typical merging DNS progenitor.
The numbers given in the figure correspond to the reference model and solar metallicity ($\Zsun$=0.02), however, the leading evolutionary channel for each model consists of the same phases.\\
Most of the merging DNS progenitors encounter stable MT and do not initiate any CE before the first SN.
The first SN in such systems is in $\sim 99\% \ [\sim 95\%]$ (for submodel $B$ [$A$])
of cases of electron-capture type.
Before the formation of the second NS the binary passes through the common envelope phase. 
We distinguish the case where it was initiated by the HG donor, as it is
uncertain whether such a CE should be formed or whether a binary could
survive at all (see section \ref{sec: assumptions:CE}).
The second SN is an iron-core collapse SN, however, in $>87 \%$ of cases the late mass transfer initiated by the expanding naked helium star occurs, which 
potentially leads to further stripping of the envelope.
Only $\sim$0.15\% out of our limited sample of ZAMS binaries (from the simulated mass range $4 \Msun <M_{a}<150 \Msun$ and $0.08 \Msun <M_{b}<150 \Msun$)
 ended up as merging DNS formed through the described channel (these binaries comprise the submodel $A$;
 $\sim$0.1\% required CE with HG donor and $\sim$0.05\% evolved through the other CE (submodel $B$)).
Also a significant number (0.01\% of these ZAMS binaries) of the merging DNS forms via accretion-induced collapse (AIC) of the WD to a NS,
yet most of them ($\gtrsim 90\%$) require CE with HG donor before the formation of the second NS and thus contribute only to submodel $A$. 
Other possible formation channels (requiring CE of any kind or no MT at all before the first SN) produce at least an order of magnitude fewer merging DNS.\\
Another variation of the main formation channel depicted in the figure \ref{fig:ref02} involves a second
episode of dynamically unstable mass transfer instead of the stable mass transfer at late stages of evolution of the binary.
This variation becomes dominant for sub-solar metallicities.
Interestingly, we find that the known properties (masses of both neutron stars, mass ratio) of the DNS associated with GW170817 within our 
simulations are best reproduced with this scenario, as presented with an example shown in figure \ref{fig:GW170817}.

\begin{figure}
	\includegraphics[width=\columnwidth]{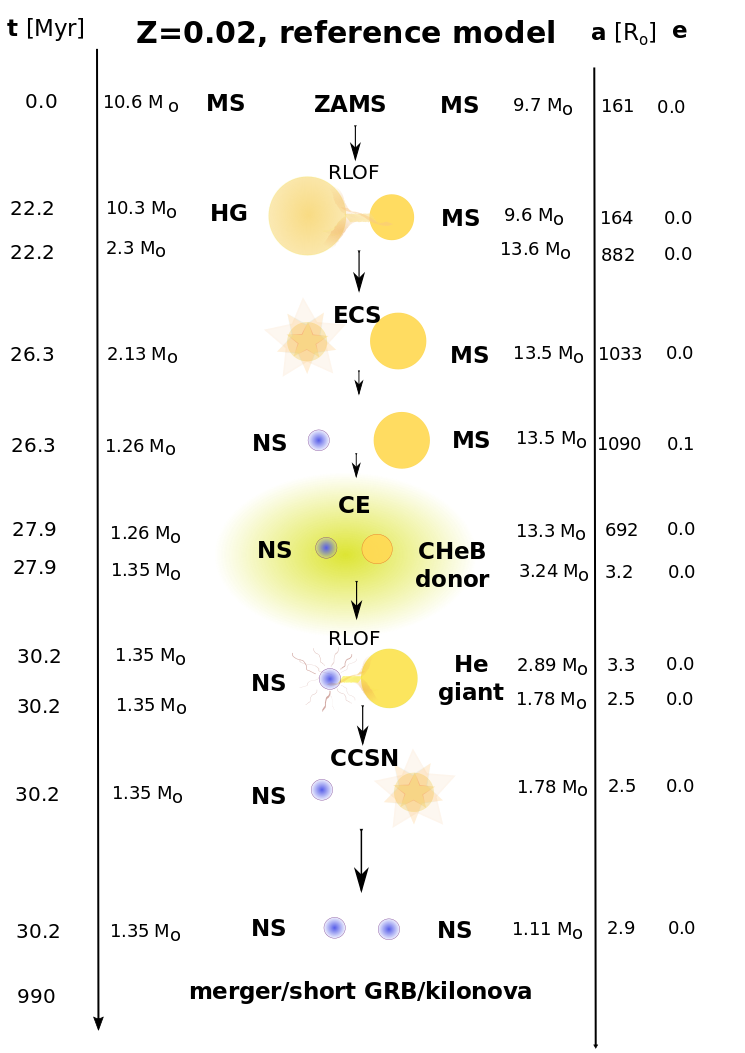}
    \caption{
      Example evolution of a typical merging double neutron star progenitor from our simulations, 
      shown for the reference model (see sec. \ref{sec: ref_model}) and solar metallicity ($\Zsun$=0.02).
      Zero age main sequence stars form most likely with relatively small masses $\sim$ 10 $\Msun$
      and pass through a stable Roche lobe overflow (RLOF) phase before the first core-collapse event.
      The first neutron star forms in an electron-capture supernova (ECS) with no natal kick.
      The binary passes through the common envelope phase (CE) initiated by core-helium burning (CHeB) star,
      shrinking the orbit and leaving behind a helium star, which later on again overfills its Roche lobe
      and initiates another phase of stable mass transfer. The second supernova occurs and a DNS system is formed,
      gradually decreasing its separation due to emission of gravitational waves
      until both stars merge, leading to a strong gravitational wave signal, short gamma-ray burst (GRB) and kilonova emission. 
    }
    \label{fig: detailed}
\end{figure}
\newpage
\subsection{Merger rates}

\begin{figure}
\begin{center}
 	\includegraphics[width=\columnwidth]{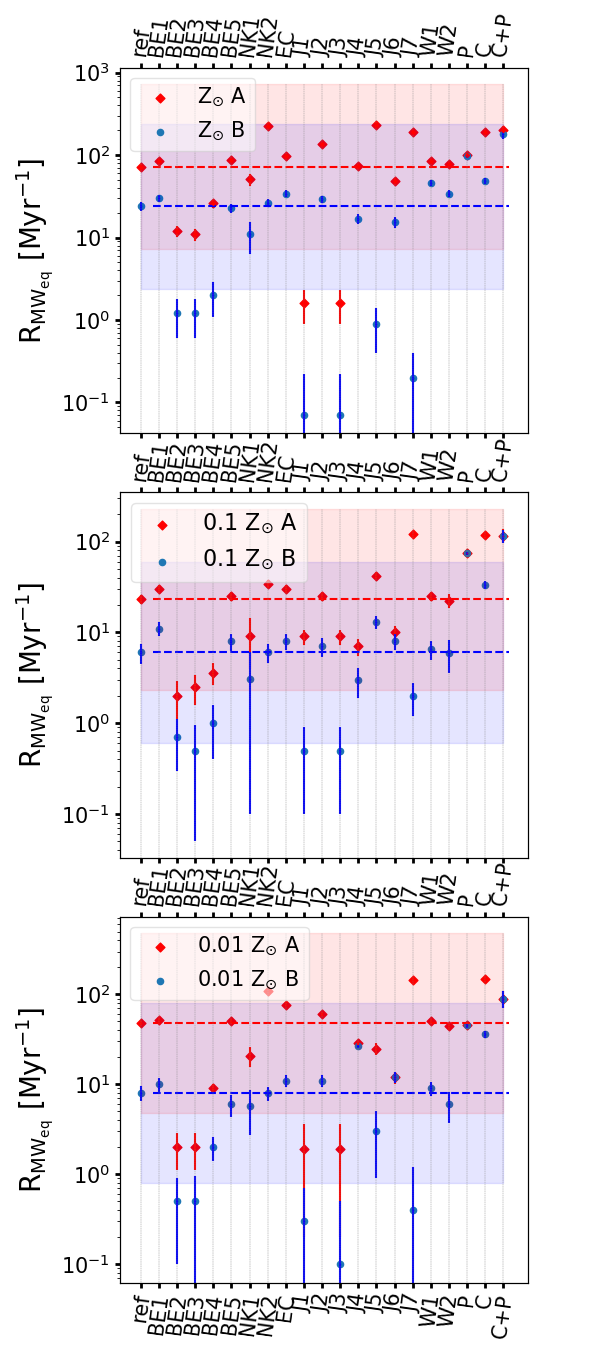}
    \caption{
    Comparison between the Galactic (Milky Way - equivalent galaxy) merger rates estimated for different models 
    (see table \ref{tab:models} for the description and table \ref{tab:results} for the numbers).
    Panels from top to bottom show the rates for solar metallicity $\Zsun$, 10\% $\Zsun$ and 1\% $\Zsun$.
    Blue circles represent the merger rates calculated without the systems undergoing CE initiated by HG donor star, while the red diamonds
    represents the estimates obtained without their exclusion. The blue (lower) and red (upper) dashed lines show the corresponding rates
    for the reference model. The vertical errorbars mark the poisson errors.
    The shaded areas around those lines limit the values located within an order of magnitude deviation from the reference model.
    }
    \label{fig:compareMWrates10}
    \end{center}
\end{figure}

The merger rates calculated for all of the models (number of merging systems, Galactic rates and local merger rate density and
approximate LIGO O3-run detection rates for models where the simulations were performed for all 32 metallicities) 
are given in the table \ref{tab:results} at the end of this paper. We focus on the differences with respect to the reference model.
Figure \ref{fig:compareMWrates10} shows the comparison between the Galactic rates obtained for different models and metallicities.
The shaded regions around the dashed horizontal lines enclose the points corresponding to the Galactic rates with the values within
an order of magnitude deviation from the reference model (blue - submodel $B$ which excludes the binaries evolving through HG-donor CE,
 red - submodel $A$ without their exclusion). 
\\
 Three immediate conclusions emerge from this comparison:
 \begin{enumerate}
 \item{ In physical models that we consider relevant for the Milky Way (submodels $B$, Z=$\Zsun$; see the top panel in fig. \ref{fig:compareMWrates10} blue),
	the typical NS-NS merger rates are moderate $\sim$ 25 Myr$^{-1}$ (ref model), with very little variation to increased rates
	(max rates: $\sim$ 100 Myr$^{-1}$: model $P$, $\sim$ 50 Myr$^{-1}$: models $W1$ and $C$), and rather significant potential to decrease
	rates (min rates: $\sim$ 0.1 Myr$^{-1}$; models $J1$, $J3$).
	None of the applied variations resulted in increase of the NS-NS merger rates with respect to the reference model
	by an order of magnitude or higher, which we would consider a significant change.}\\
 
 \item{ In many cases the difference between the submodel $A$ and $B$ within the same model is bigger than an order of magnitude,
      especially in cases with increased angular momentum losses in the $J$-set of models,
      but also in the models $BE2$, $BE3$ and $NK2$ with non-standard treatment of the natal kicks.
      An extreme example is the model $J7$ where this difference for the solar metallicity is $\sim 10^{3}$.}\\
 
 \item{ Certain variations applied to the standard treatment of the angular momentum loss and the natal kicks can 
      severely decrease the predicted merger rates for NS-NS binaries.
      The biggest drop with respect to the reference model (over two orders of magnitude) is present at $\Zsun$ in submodel $B$ for models $J1$ and $J3$ with extremely
      high angular momentum loss during the stable MT with non-degenerate accretor. 
      The change in other direction: models $J2$ and $J4$ with very low  angular momentum losses, does not reveal a similarly considerable effect on the merger rates
     }
 \end{enumerate}

 \subsubsection{The reference-like models}
There is a group of models for which the Galactic merger rates in all cases, i.e. for all metallicities and in both submodel $A$ and $B$
remain close (within a factor of 3) to the reference model. These are $BE1$, $BE5$, $NK1$, $EC$, $J2$, $J4$, $W1$ and $W2$.
We refer to them as the reference-like models.
The rates predicted for model $J6$ (with lowered angular momentum losses during MT with NS accretor) always stay close to the reference value in
variation $B$, while for the submodel $A$ the rates drop for sub-solar metallicities with respect to the submodel $A$ reference value, 
so that variations $A$ and $B$ lead to the same results (no CE initiated by HG stars).
We consider these deviations from the reference model as small (changes are comparable to the effect of the use of a different normalization)
and will not discuss these models in more details.

\subsubsection{Models leading to increased rates}
We find just two models in which the rates are systematically higher than the reference value by more than a factor of 2 in all cases and it is the model $C$ which 
combines several factors favoring the formation of the merging DNS (combination of the reference-like models $EC$, $J2$ and $BE1$) and another version of this model
called $C+P$, which incorporates also variation introduced in model $P$ (discussed in sec. \ref{sec: P}).
The decreased angular momentum loss during the MT with non-degenerate accretor ($J2$) pushes more systems to the preferred
formation path (see fig. \ref{fig:ref02}) of the merging DNS, by avoiding CE before the formation of the first compact remnant.
The orbit in this case stays relatively wider as the primary star expands and the early interaction is more likely to remain stable.\\
The wider range for the core mass which can lead to an electron-capture triggered collapse ($EC$) allows more stars to explode
in ECS during the formation of the first NS, decreasing the fraction of systems disrupting during the critical first supernova (as there is no
natal kick added). The inclusion of $BE1$ variation additionally increases the number of binaries surviving the second SN. Most of the potential
progenitors of the merging DNS are stripped of their outer layers during the late mass transfer phases, thus when the natal kick velocity
is assumed to be proportional to the mass of the ejected envelope, the second NS generally forms with smaller birth velocity than in the
reference model. 
\\
Also relatively high rates (irrespective of the metallicity) are predicted for the submodel $J7\ A$, in which the angular momentum losses during MT with NS accretor were doubled.
In this case all binaries consisting of a NS and an evolving companion that in the reference model would pass through CE and then encounter the stable mass transfer phase
initiated by an expanding helium star as the core helium burning ceases ('helium Hertzsprung gap'), encounter the second CE instead, which is much more efficient in shrinking the orbit,
and thus increasing the number of merging systems. As the donor is a HG star, these binaries always add up to submodel $A$, as a consequence lowering the rates for the $B$ variation in this model.
Similarly in submodel $J5 A$ (fully-conservative mass transfer) for $\Zsun$ the rates grow due to increased number of systems encountering two subsequent CE phases after the formation
of the first NS. However this time it is due to the fact that the secondary star gained more mass while still being on the main sequence 
(as the whole transferred material is added to the secondary star) during the stable MT. 
It affects its later evolution so that it initiates the CE during the core helium burning phase rather than on the red giant branch, which leads
to smaller final separation.
This is due to the different value of $\lambda$ parameter (describing the binding energy of the envelope) at these stages 
- the value is smaller during the core helium burning, so more orbital energy is needed to eject the envelope than in case of CE with a donor star on the red giant branch. 
Smaller separation at the end of this common envelope evolution favors the occurrence of the subsequent CE phase with HG donor \citep[see][for more thorough discussion of this case]{Dominik12}.\\
Model $NK2$, which forces half of the CCSN to lead to small natal kicks ($<$ 50 km/s) allows to increase the rates only in submodel $A$, while for $B$ the rates remain
very close to the reference values. In this variation twice more binaries with NS survive the first SN. These 'additional' systems are relatively wide and otherwise would be
disrupted, or even more widened due to large natal kick, so that they would not interact anymore and not contribute to the merging population.
The binaries that received a chance to survive the first SN in this model need to decrease their orbital separation even more than those that would survive under the standard conditions.
We find that in order to contribute to the merging population, they have to evolve through the two subsequent common envelope phases as in the model $J5 \ A$ and so they increase
the number of the merging systems only if the common envelope with the HG donor is allowed.

\subsubsection{Models leading to decreased rates}

\begin{figure}
\begin{center}
 	\includegraphics[scale=0.55]{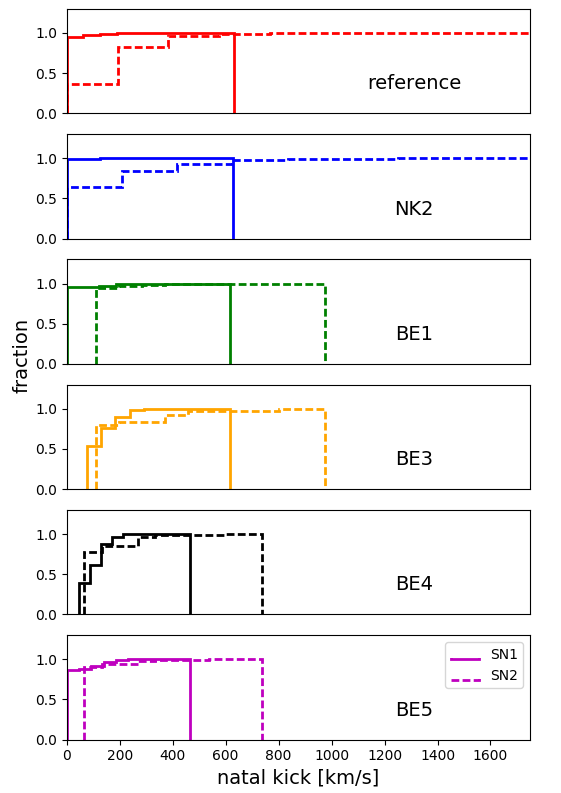}
    \caption{
    Comparison of the magnitudes of the natal kick velocities gained by NS that form the merging double neutron stars.
    The solid line shows the cumulative distribution of the natal kicks gained after the first SN in the binary,
    while the dashed one demonstrates the cumulative distribution of these velocities after the second SN.
    The distribution for model $BE2$ is nearly identical to the one presented for the model $BE3$.
    As before the formation of the first compact object the orbit is still relatively wide, the first NS in the progenitor
    of merging double neutron star system forms with very small natal kick velocity (otherwise the binary is disrupted;
    note that these velocities can be much bigger after the second SN, when the orbit is much more compact).
    }
    \label{fig:compareKicks}
    \end{center}
\end{figure}

Similarly, one can distinguish the low-rate group of models which in all cases lead to at least twice lower estimates than the reference model.
These are $BE2$, $BE3$, $BE4$ - models with the natal kicks treated according to the prescription suggested by \citet{BrayEldridge16}, but applied
to a NS that formed in CCSN, ECS and in $BE4,3$ also AIC, and also the models $J1$ and $J3$ with high angular momentum losses during MT with non-compact accretor.
In the listed $BE$-variations the decrease is a consequence of adding a natal kick velocity after the NS formation in an ECS.
As can be seen in the figure \ref{fig:compareKicks}, in those models which allow for it, the first NS forms with the natal kick close to zero.
In the reference model the first SN in 99\% (in submodel $B$, 95\% in submodel $A$) of the merging DNS binaries was of electron-capture type, where the zero natal kick guaranteed that the
binary will remain bound. 
In $BE4$($BE2,3$) the natal kick after ECS is at least 70 km/s (120km/s) and can be higher depending on the mass of the ejected envelope 
(in our simulations this mass is within the range 0.12-0.9 $\Msun$).
This causes many potential progenitors of the merging NS-NS binaries to disrupt and hence lowers the rates.
Similarly, the drop in the number of merging DNS (albeit smaller) is also present in model $NK1$, where the natal kicks after ECS are drawn from a 
Maxwellian distribution with 1-D $\sigma = 26.5 \rm  \ km/s$ peaking around $40$ km/s.
In models $J1$ and $J3$ the extreme angular momentum loss during the early mass transfer practically quenches the formation channel described in \ref{ap: evolution}.
The merging DNS progenitors pass through the common envelope before the first NS can form, where the great majority of them coalesces prematurely.\\
The rates in submodels $J5 \ B$ and $J7 \ B$ drop due to the reasons discussed in the previous section.
 
\newpage
 
 \subsection{Model P} \label{sec: P}

In variation $P$ we force a thermal-timescale stable mass transfer in place of the common envelope initiated by a HG donor star, when the accreting star is a NS or BH.
This is the very provisional attempt to demonstrate the effect of the recent findings of \citet{PavlovskiiIvanova15} and \citet{Pavlovskii17} discussed in
\ref{sec: assumptions:CE} and should be treated with caution,
as the case of the NS-accretor was not studied in detail by the authors and neither the timescale of the stable mass transfer, 
nor the exact condition for the eventual onset of the unstable mass transfer are known.
Our forced MT, if not completed earlier, may instantaneously finish with the common envelope when the donor star evolves off the Hertzsprung-gap, 
if the standard conditions for the onset of CE are then fulfilled.
However, at $\Zsun$ nearly half of the merging population forms with no CE at all and this fraction is even higher for 10\% of the solar metallicity.
The MT with a compact accretor is limited by the Eddington rate and if the rate is exceeded, the mass is lost from the binary (non-conservative MT; in the discussed case the MT
is highly non-conservative as the Eddington rate is exceeded by a few orders of magnitude) and the excess material escapes from the vicinity of the compact accretor,
carrying away significant angular momentum ($j_{loss}\sim M_{\rm donor}/ M_{\rm accretor}$).
Such a mass transfer with NS always leads to decrease in the orbital separation ($M_{\rm donor}>M_{\rm accretor}$),
favoring the formation of the merging system.
In fact, we find that the merger rates increased with respect to the reference model and variation $A$ and $B$ lead to nearly the same result, differing only
(there still can be a CE with HG donor with non-compact/WD accretor), due to CE initiated by HG donor with WD in the AIC channel.

\section{Confrontation with the observational limits}

\subsection{Observational limits for DNS merger rates}\label{sec: obs_rates}

There are several astrophysical sources on which the observational limits on DNS coalescence rates can base. 
The first approach benefits from the fact that, unlike in case of BHNS or BHBH binaries (apart from the merger signatures),
we directly observe DNS systems and relies on the investigation of the properties of the Galactic population of double neutron 
stars \citep*{Kim10,OShaughnessy_Kim10,Kim15}.
By properly including the selection effects of the pulsar searches which detected these systems, one can estimate the total rate 
of merging DNS binaries within the galaxy.
To date there are 12 known DNS residing within the Galactic disc, one 
candidate system\footnote{http://www.atnf.csiro.au/ATNF-DailyImage/archive/2017/25-Jan-2017.html}  and two 
binaries found within globular clusters \citep[see table 1 in][and references therein]{Chruslinska17},
where the dynamical formation scenario cannot be excluded and thus they cannot be used to constrain the merger rates for the isolated channel.
The most recent estimates that use the disc Galactic sample were obtained by \citet{Kim15} and are based on three systems B1913+16, B1534+12,
and J0737-3039A,B (the double pulsar, both pulsar A and B were used to constrain the rates). 
The authors provide the range of 7 - 49 $\rm Myr^{-1}$ at 95 percent confidence, with the median value of 21 $\rm Myr^{-1}$ 
as their best estimate for the NS-NS Galactic merger rates. 
However, this type of analysis has a number of important caveats.
First, the number of observed merging DNS is very scarce and inclusion of a \emph{single} newly observed system may affect the predicted
rates by a factor of a few as revealed by the example of J1906+0746 provided by \citet{Kim10} which, if included in the analysis,
would increase the rates by a factor of $\sim$ 2. To our knowledge, these estimates were not revised to evaluate the impact of the most recently
discovered systems.
One should also bear in mind that such extrapolations assume that the observed sample is representative of the underlying Galactic population.
Uncertainties may also arise due to the estimation of the beaming fraction (the fraction of the sky covered by the pulsar radiation beam)
and the pulsar age \citep{OShaughnessy_Kim10}.
However, probably the biggest uncertainties are related to the treatment of the pulsar luminosity function, necessary
to model the selection effects, but weakly constrained by observations \citep[e.g. appendix in][]{OShaughnessy_Kim10}.
In fact, different choices for the pulsar luminosity distribution are allowed within the observational limits and
these can change the estimated merger rates by an order of magnitude \citep{Mandel_OShaughnessy10}.
\\
Apart from the Galactic population short gamma ray bursts (sGRB) may provide constraints on DNS merger rates.
The binary compact object merger scenario is so far the most successful in explaining the observational properties of short GRB.
However, NS-NS mergers are only one of the possible sGRB progenitors and currently other candidates (BH-NS mergers)
cannot be excluded \citep[e.g.][]{Berger14}. As a result, they may contribute to only a fraction of the observed events.
\citet*{Petrillo13} estimated the merger rates based on the sGRB observed by SWIFT,
restricting the analyzed sample to observations with well determined redshift and certain association to an optical counterpart.
They obtained the coalescence rates in the local universe ranging from 500 to 1500 $\rm Gpc^{-3} \ yr^{-1}$.
This result is, however, strongly dependent on the weakly constrained beaming angle $\theta_{j}$ of the collimated emission from
the short gamma ray bursts, as stressed by the authors (see figure 3 therein). 
Increasing the angle would relax the limits on merger rates. If the angle is a factor of two
smaller than assumed by the authors, the merger rate could be as high as several thousand $\rm Gpc^{-3} \ yr^{-1}$ ($\theta_{j} \sim 10^{\circ}$), 
while increasing the angle by a factor of two would decrease the rate to around 200 $\rm Gpc^{-3} \ yr^{-1}$ ($\theta_{j} \sim 40^{\circ}$).
\\
The question of the beaming angle and its impact on the merger rates that base on sGRB was also addressed by \citet{Fong15}.
They show that using 4 events for which the measurement of $\theta_{j}$ was possible would lead to the median $\theta_{j}=6\pm1^{\circ}$ for sGRB. 
However, present observations are biased towards the events with small $\theta_{j}$, as
most of the events fade before the detection of the jet break (necessary for determination of the opening angle) is possible
and only allow for a lower limit on $\theta_{j}$.
The authors include 7 events with the lower limits $\theta_{j} \gtrsim 5^{\circ}$ (arguing that shallower lower limits 
would not have a significant effect on the $\theta_{j}$ distribution) and show that if the whole range of opening angles is allowed
($\theta_{j}\leqslant 90^{\circ}$) the median would shift to $\theta_{j}=33_{-27^{\circ}}^{+38^{\circ}}$.
Setting the maximum opening angle $\theta_{j;max}=30^{\circ}$ \citep[as suggested by the simulations, e.g.][]{Rezzolla11} 
results in the median of $\theta_{j}=16_{-10^{\circ}}^{+11^{\circ}}$. Using this last estimate \citet{Fong15}
obtain the local merger rate density of $270^{+1580}_{-180} \ \rm Gpc^{-3} \ yr^{-1}$.
\\
Another estimate is based on the analysis of a sample of 8 sGRB observed by SWIFT with well determined redshifts and comes from \citet{Coward12}.
They infer the rates by relying on single events and correcting for biases of the detector, avoiding the uncertain distributions 
needed to describe the whole population (e.g. beaming angles, luminosity function).
They provide the limits for the local sGRB rate density for two extreme cases: the lower limit assuming isotropic emission 
and the upper limit using the observed beaming angle of each sGRB or the smallest $\theta_{j}$ from the data if the measurement is unavailable.
The obtained values are $8^{+5}_{-3}$ $\rm Gpc^{-3} \ yr^{-1}$ and  $1100^{+700}_{-470}$  $\rm Gpc^{-3} \ yr^{-1}$ respectively.
\\
Theoretically one expects a weak optical/nIR afterglow associated with sGRB, due to radioactive decay of the r-process material
produced during the NS-NS/BH-NS merger, so called kilonova (or macronova) \citep{LiPaczynski98}.
Besides the one associated with the first detection of gravitational waves from merging DNS, 
so far two observational candidates for kilonovae were reported \citep{Berger13,Tanvir13,Jin16}.
Using the latter two observations as constraints \citet{Jin16} estimated the local compact objects merger rate 
(NS-NS and/or BH-NS mergers) to be $\rm 16.3^{+16.3}_{-8.2} \ Gpc^{-3} yr^{-1}$ (dependent on the beaming angle).
The authors stress, however, that this estimate should be taken as a lower limit.
This estimate of course suffers from the very poor number of constraining observations and will likely change when
the new observation will be taken into account.
Furthermore, the interpretation of these observations as kilonovae was questioned with the recent simulation results of the emission from the
r-process ejecta \citep{Fontes17}.
We thus quote the above results for completeness, but focus mainly on the other observational constraints.
\\
Finally, the first detection of gravitational waves from merging double neutron stars allows to put
another constraint on the local merger rate density of these systems \citep{1stNSNS_merger_paper}, 
resulting in the range of $1540^{+3200}_{-1220}$ $\rm Gpc^{-3} \ yr^{-1}$.

\subsection{Comparison with observations}

\subsubsection{The Galactic rates}
\begin{figure}
\begin{center}
 	\includegraphics[scale=0.45]{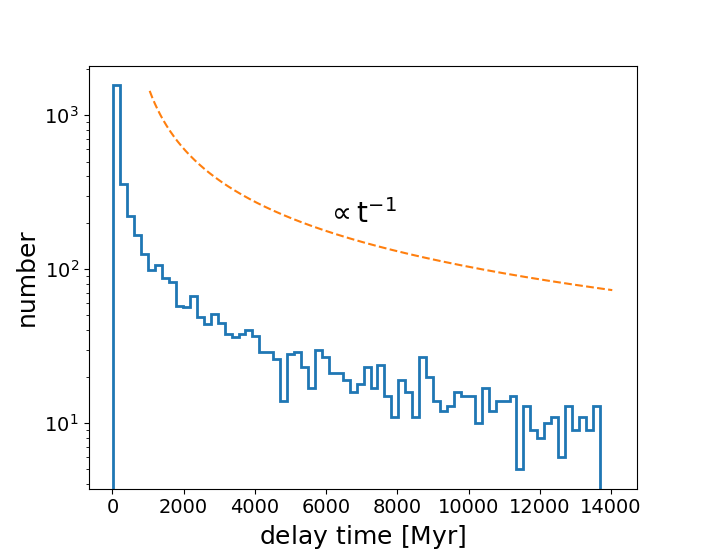}
    \caption{
    Histogram of the delay times plotted for the reference model and solar metallicity. The line $\propto t^{-1}$ is shown for comparison.
    The distribution of the delay times is strongly peaked at a few hundreds Myr and then decreases sharply irrespective of the model.
    }
    \label{fig: Tdel}
    \end{center}
\end{figure}

We begin with a comparison of our Galactic merger rates with the limits imposed by the observations of the Galactic population
of merging DNS discussed in \ref{sec: obs_rates}. 
As the distribution of the delay times (the difference between the merger time and the birth time; see fig. \ref{fig: Tdel}) 
for merging DNS in our simulations is
strongly peaked at a few hundred Myr and then decreases sharply irrespective of the model, the present-day merging population will 
be composed mainly of the binaries that formed a few hundred Myr ago. The observational studies of the age-metallicity relation
for the Milky Way reveal that this relation is nearly flat until $\sim$8 Gyr ago \citep[e.g.][]{Bergemann14}.
Although a significant scatter in metallicity around $\Zsun$ is found at any age, 
it drops below 30\% $\Zsun$ only at look-back times greater than 8 Gyr.

\begin{figure*}
\begin{center}
 	\includegraphics[scale=0.7]{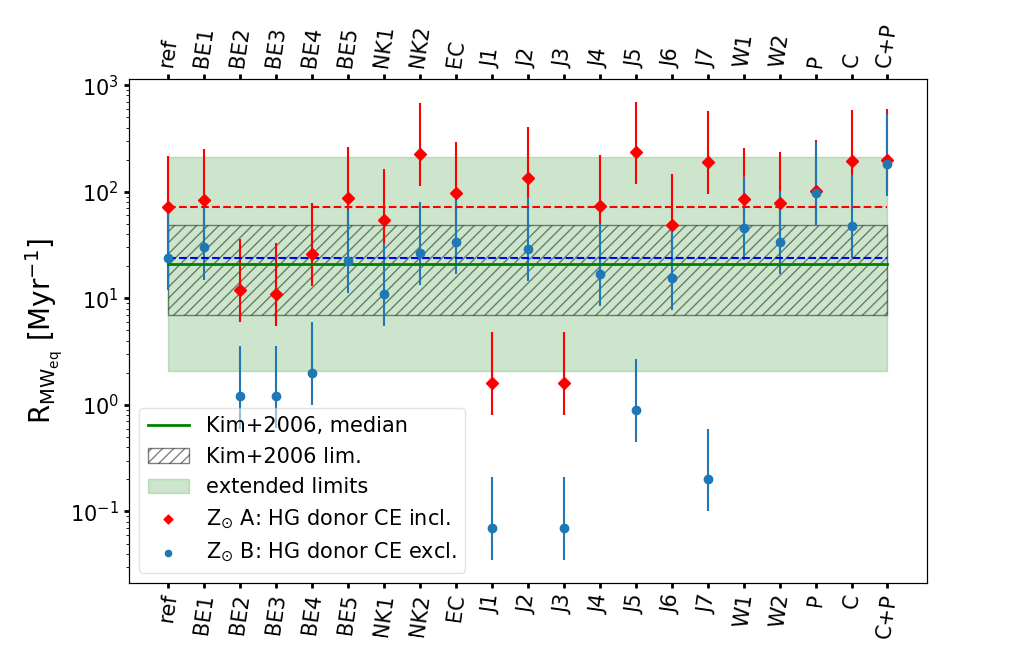}
    \caption{
     Comparison between the Galactic DNS merger rates estimated for different population synthesis models (see tables \ref{tab:models} and
    \ref{tab:results}) and the observational limits obtained by \citet{Kim15}.
    The red diamonds show the rates obtained for submodel $A$, while blue dots for submodel $B$, the red and blue dashed lines mark 
    the corresponding values obtained for the reference model. The vertical bars show the spread by a factor of two up and down around
    the estimated merger rate values. The green line marks the best estimate obtained by Kim et al., while the hatched region indicates
    the reported 95 percent confidence limits around the median value.
    The shaded region extends to a factor of 10 up and down around this median, corresponding to the uncertainties due to different assumptions
    about the pulsar luminosity distribution.
    }
    \label{fig: obs_galactic_rates}
    \end{center}
\end{figure*}

We thus use the Galactic rates obtained for solar metallicity for comparison with the results obtained by \citet{Kim15}.
This comparison is depicted in figure \ref{fig: obs_galactic_rates}. The simulations results are shown as red diamonds (submodel A) 
and blue dots (submodel B) with the dashed lines in the corresponding colours marking the values obtained for the reference model.
The vertical 'errorbars' spreading by a factor of 2 up and down were added to the estimated values, as a conservative limits
on the systematic changes (affecting all of the models) of the results due to different calculation of the normalization 
(e.g. the higher constant Galactic SFR as used by \citet{MennekensVanbeveren14} would increase all of the rates by a factor of $\sim$1.3)
or different assumptions about the binary fraction (e.g. using low binary fraction $f_{\rm bin}$=0.5 as in \citet{Dominik12} 
would decrease the rates by a factor of $\sim$1.8).
The green horizontal line marks the location of the median value obtained by \citet{Kim15}
and the hatched region the uncertainties to their best estimate. Additionally, the shaded green region encloses the merger rate estimates
contained within an order of magnitude deviation from the median found by Kim et al., to indicate the potential uncertainties due to
different assumptions about the pulsar luminosity distribution (we will refer to these limits as extended limits, to distinguish from the
more restrictive bounds provided by \citet{Kim15} based on their reference model).
\\
It is found that the Galactic merger rate obtained for our reference model (24 Myr$^{-1}$) corresponds almost exactly to the
median of the most recent DNS merger rate estimate for the Galaxy \citep[21 Myr$^{-1}$][]{Kim15}.
Thus, the $B$ submodels of all of the reference-like models are consistent with the observations within the more restrictive limits
(the hatched region). 
These contain also model $J6$ and submodels $BE2,3,4 \ A$ and $NK2 \ B$.
The reference value for the submodel $A$ (and most of the rates for other models in this variation) lies outside these restrictive bounds, however,
it cannot be excluded as long as the extended limits are considered.
Only submodels $J1,3,7 \ B$ with extreme angular momentum losses lie completely outside the extended observational limits.

\subsubsection{The local merger rate density}

\begin{figure*}
\begin{center}
 	\includegraphics[scale=0.65]{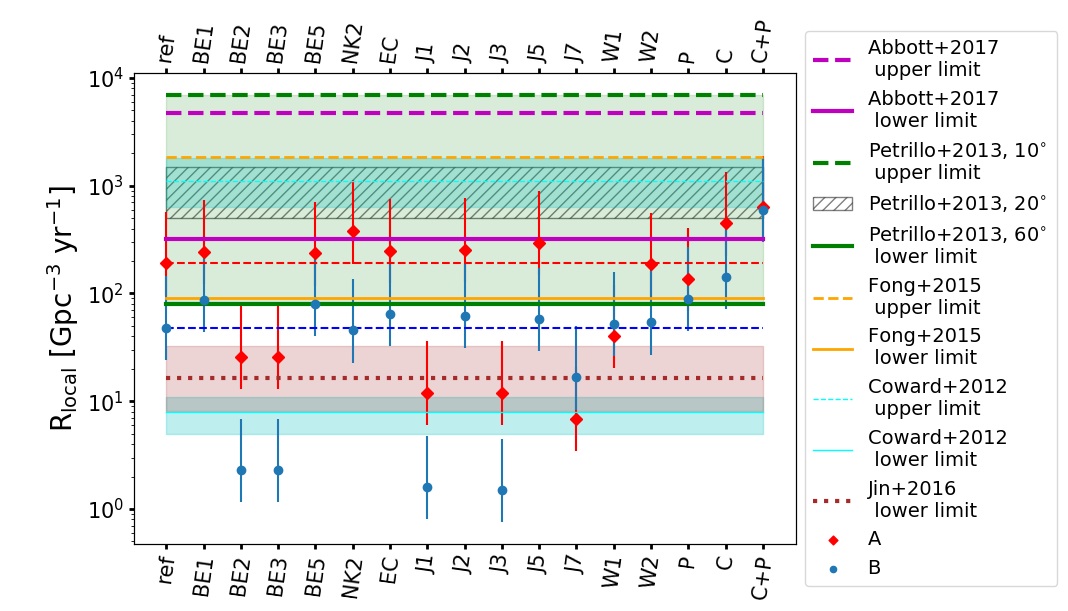}
    \caption{
    Comparison between the local DNS merger rate density estimated for different population synthesis models (see tables \ref{tab:models} and
    \ref{tab:results}) and the observational limits on short GRBs and DNS merger rates.
    The red diamonds show the rates obtained from the simulations for variation $A$ (HG donors in CE allowed),
    while blue dots for variation $B$ (HG donors in CE not allowed).
    The red and blue dashed lines mark the corresponding values obtained for the reference model (ref). 
    The vertical bars show the spread by a factor of two up and down around the estimated merger rate values.
    The purple lines mark the upper (dashed) and lower (solid) limits obtained from the first detection of gravitational waves
    from the merging double neutron star \citep{1stNSNS_merger_paper}.
    The hatched region indicates the range  obtained by \citet{Petrillo13} based on the short GRB observations, with the default value of the
    beaming angle (20$^{\circ}$). The green shaded region spans between the upper limit assuming the beaming angle 10$^{\circ}$
    and the lower limit assuming weakly collimated emission (60$^{\circ}$, see figure 3 in Petrillo et al.).
    The orange lines mark the upper (dashed) and lower (solid) limits reported by \citet{Fong15}, also based on sGRB observations.
    The limits from another sGRB study by \citet{Coward12} are plotted in light blue. 
    The shaded regions in the same colour indicate the uncertainties of these calculations reported by the authors.
    The brown line indicates the lower limit reported by \citet{Jin16} based on the observations of the potential two kilonova candidates
    (the shaded region gives the uncertainty limits; kilonova associated with GW170817 is not included and will likely affect the range).}
    \label{fig: obs_local_rates}
    \end{center}
\end{figure*}

We provide the estimates of the local merger rates calculated as described in \ref{ap: rates} for 17 out of 21 models presented in this study 
(for merely technical reasons, as it is computationally expensive to obtain the simulations for the full set of 32 metallicities which are used to
calculate the rates).
For those models we confront the calculated rates with the observational limits from sGRB observations and gravitational waves searches.
The comparison is depicted in figure \ref{fig: obs_local_rates}.
There is a number of poorly constrained components within our
cosmological calculations that are responsible for the buildup of
the final uncertainty associated with our estimated NS-NS merger rate density (see appendix \ref{ap: rate_err}).
These uncertainties are still much smaller than those associated with the evolutionary assumptions, reflected in
differences between various models presented in this study.
Similarly as in the case of the Galactic rates, we put a rough limit on these uncertainties,
 allowing for a variation within a factor of 2 of our estimates.
However, a more careful evaluation of the uncertainties related to such calculations would be desirable.
The rates for submodel $A$ and $B$ are plotted as in figure \ref{fig: obs_galactic_rates}.
We show the limits based on the first detection of gravitational waves from DNS merger \citep{1stNSNS_merger_paper} (purple lines).
We then indicate the short GRB based limits on the DNS merger rate density as discussed in sec. \ref{sec: obs_rates}.
The hatched region shows the limits derived by \citet{Petrillo13} based on sGRB observations,
using their standard value of the beaming angle of 20$^{\circ}$.
Green lines indicate the upper limit estimated by these authors if this angle would be smaller ($\sim$10$^{\circ}$) and
the lower limit obtained assuming a weakly collimated emission (60$^{\circ}$).
The orange lines correspond to the limits provided by \citet{Fong15} and 
the light blue lines show the upper and lower limits found by \citet{Coward12}.
Finally, the lower limit reported by \citet{Jin16} based on the observations of the potential two kilonova candidates
is marked with the brown line.
\\ \newline
Only three of the models probed within this study produce enough NS-NS mergers in the local Universe to meet
the most constraining observational limits on the local DNS merger rate density that were imposed by the first
 detection of gravitational waves from coalescing double neutron stars.
These are our combined models: $C$ in variation $A$ and $C+P$ (differing in the treatment of unstable mass transfer with HG donor and NS/BH 
accretor), as well as submodel $NK2 \ A$ which introduces a high fraction of CCSN leading to small ($\leqslant$50 km/s) natal kicks.
A number of other models in variation $A$ is only marginally (within the uncertainty limits) consistent with this observational constraint.
These are variations $A$ of the models $BE1,5$, $ECS$, $J2,5$, $W2$, $P$ and our reference model.
The other models, including the most standard one - reference in variation $B$, fail to meet the constraints based on gravitational waves observations.
\\
The last column of table \ref{tab:results} contains the approximate LIGO detection rates 
(number of expected detections per year of observations),
calculated using the formula \ref{eq: detection} and assuming the detector will reach the 
maximum expected O3-run DNS range and its intended DNS range for the advanced stage 
\citep[corresponding to $\rm D_{h} \sim 385$ Mpc and $\sim$487 Mpc respectively][]{Abbott16b}.
The three models that fall within the limits from \citet{1stNSNS_merger_paper} predict $\sim$8-12 detections per year
of operation of the advanced detectors with their full planned sensitivity. 
\newline
\\
All of the models ($A$ and $B$, all variations) fall below the upper limits on sGRBs calculated by Petrilo et al., Fong et al. and Coward et al.
Our physical models $B$, including the reference model are mostly below or very close to the sGRB lower limits reported
by Fong et al. and Petrillo et al. (which account for the beamed emission from sGRB, in contrast to the very broad lower limit
by Coward et al., calculated for isotropic emission).
This indicates that our model framework $B$ that seems to well reproduce DNS merger rates in the Milky Way
is not sufficient to deliver enough mergers to explain all cosmic sGRB. 
However, it should be noted that NS-NS mergers are only one of the potential progenitors of sGRB
and some fraction of these events may be associated with BH-NS mergers, so it is presently not justifiable to rule out these models
on that basis.
Most of the models in variation $A$ (apart from $BE2,3$, $W1$ and $J1,3,7$) are above these lower limits.
Submodel $A$ is found to over-produce BH-BH mergers \citep{Belczynski16N}, however, it cannot be ruled out 
for lower mass stars (DNS progenitors).\\
Submodels $BE2,3 \ B$ (requiring substantial natal kicks after electron-capture SN) and $J1,3 \ B$, $J7 \ A$ (assuming extremely high angular momentum loss during the mass transfer) 
leading to the lowest estimated rates fall below the lower limit from the study of \citet{Jin16}.
Submodels $BE2,3 \ B$ are marginally consistent with the broad lower limit reported by \citet{Coward12},
while $J1,3 \ B$ are in tension with this estimate.
These models can be excluded as non-physical within the framework of our calculations.

\section{Discussion}

We find that the treatment of the common envelope remains the main source of evolutionary uncertainties,
affecting the predictions of the merger rates for double neutron stars by even more than two orders of magnitude (e.g. our submodel $J7 \ A$ and $J7 \ B$).
The importance of this phase for the merger rates was analyzed by a number of authors, who focused on different assumptions regarding the CE evolution.
For instance, \citet{VossTauris03} investigated the effect of changing the critical mass ratio required for the mass transfer with helium stars to become unstable.
They find that the changes of this ratio can decrease the DNS rates by a factor of $\sim$3. 
Several studies demonstrate the effect of using different, fixed values (although it is generally accepted that this 
parameter should vary depending on the evolutionary stage of the donor)
 of the envelope binding energy parameter $\lambda$ \citep[e.g.][]{VossTauris03,Dominik12}. 
 Using $\lambda$=0.5 \citet{VossTauris03} find the increase in the rates with respect to their reference model, while \citet{Dominik12} find that either using
 very low or very high values decreases the rates. The codes used in both studies use different prescriptions for the parameters of the CE in their reference
 models, thus the comparison is not straightforward, nonetheless both identify $\lambda$ as a factor that can easily shift the DNS rates
 by an order of magnitude.
 \citet{Dominik12} and \citet{MennekensVanbeveren14} show the influence of varying the second parameter of the CE $\alpha_{\rm CE}$ (efficiency of the ejection of the 
 envelope when decreasing the orbital separation).
 Both groups show that increasing the $\alpha_{\rm CE}$ 5 times can affect the rates by a factor of a few
 (\citet{Dominik12} find the increase in submodel $A$ and decrease in $B$, while \citet{MennekensVanbeveren14} find the rates to increase).
 Lowering $\alpha_{\rm CE}$ decreases the rates in \citet{MennekensVanbeveren14} (lowering $\alpha_{\rm CE}$ 10 times makes DNS merger rates drop to zero; see their model 6 and 8),
 while \citet{Dominik12} find a slight increase (in both variation $A$ and $B$) when decreasing $\alpha_{\rm CE}$ 5 times.\\
 Here, motivated by the recent studies by \citet{PavlovskiiIvanova15} and \citet{Pavlovskii17} we pay particular attention to the case when CE is initiated by the donor star
 passing through the Hertzsprung gap phase. Apart from differentiating between the cases when we allow for the CE in that case to ensue ($A$) and the case when we
 exclude those systems ($B$), assuming that they should not contribute to the merging population, we also include a model $P$ where we force the stable
 thermal-timescale mass transfer in place of the CE initiated by a HG star when the companion is a NS or a BH. 
 The results show that the simple exclusion of these systems may not be justified. The stable mass transfer with a compact accretor does not lead to 
 a net expansion of the orbital separation as is often the case with an early MT phase and in fact is able to shrink the orbit enough to allow for the 
 formation of a merging NS-NS system, increasing the Galactic rates by a factor of 4-12 (depending on the metallicity) with respect to the reference model,
 when submodels $B$ are compared and only slightly affecting the rates for submodel $A$.
 We stress that our model is a very crude attempt to address the results of Pavlovskii et al.,
 however it shows the importance of further investigation of the Roche lobe overflow events initiated by stars with no steep density contrast between
the core and the envelope (as HG stars), from both the side of the mass transfer and common envelope dynamics and from the point of view of the binary evolution
and properties of the merging populations. 
On top of this, in many cases we find that within the same model the Galactic rates predicted in variation $A$ and $B$ differ 
by more than an order of magnitude, especially when the assumptions about the angular momentum loss are varied, which is comparable to the largest deviation
in the estimated merger rates found between different models studied in this work.
\\ \newline
The predicted merger rates are also known to be sensitive to the assumptions related to the natal kicks. 
Almost every parameter study performed within the population synthesis scheme contains a variation allowing for non-standard (usually smaller velocities drawn
from the Maxwellian-like distribution shifted towards lower kicks) 
magnitudes of the natal kicks gained by the supernova remnants \citep[e.g.][]{VossTauris03,Bogomazov2006,Dominik12,MennekensVanbeveren14},
as generally apart from the observational requirements that some neutron stars must be able to form with large velocities (even in excess of 1000 km/s as revealed by pulsar
proper motions) and some must form with small or none natal kicks (to remain bound within a binary or retained within a globular cluster), 
they constitute another weakly constrained ingredient of the theory of binary evolution.
Lowering the natal kicks increases the fraction of binaries that can survive the supernova explosion and thus allows more systems to reach the stage of the DNS.
This is generally found to increase the rates by a factor of a few 
\citep[e.g.][except for \citet{Bogomazov2006} who find the increase even greater than an order of magnitude, however, when they allow for the
formation in a standard CCSN scenario with very low or even zero natal kicks]{VossTauris03,Dominik12,MennekensVanbeveren14,Chruslinska17}.
The increase in the rates is not spectacular even in model $NK2$ where we allow the half of the CCSN to lead to small natal kicks ($\leqslant 50$ km/s chosen from a flat distribution),
and present only in submodel $A$. It helps to avoid disruption, however, the additional binaries (surviving due to this modification) are wide and unless
this modification is paired with the enhanced mechanism which will act to decrease the separation, the rates will not grow more than by a factor of a few.
\citet{MennekensVanbeveren14} and \citet{Chruslinska17} also distinguish the formation channel involving ECS/AIC and modify the assumption
in which it leads to no natal kick. \citet{MennekensVanbeveren14} find the decrease in DNS merger rates by a factor $\sim$7 when they 'switch off' the ECS channel,
which probably means that ECS-formed NS receives a standard high kick. The variation from \citet{Chruslinska17} is also repeated here (model $NK1$). In this
model ECS-formed NS receives a plausible small natal kick $\sim 50$ km/s and the corresponding change in the rates
with respect to the reference model is small - within a factor of two (even smaller in submodel $A$).
\\
The significant drop in the rates is present in models where we used the prescription proposed by \citet{BrayEldridge16}, in which the magnitude of the natal kick
is proportional to the ratio of the masses of the ejected envelope and formed compact remnant, applying it to both CCSN and ECS(and AIC in some variations, although
the differences between the models affecting both ECS and AIC and those affecting only ECS are marginal).
The Galactic rates for those models when we do not allow for CE with HG donors ($BE2,3,4 \ B$) are only marginally consistent with the extended observational limits
based on the properties of the sample of Galactic DNS. 
The local merger rate density calculated for the models $BE2,3 \ B$ is in tension with the observational limits based on sGRB and potential kilonovae observations.
The natal kicks received by ECS-formed NS in these variations are generally bigger than expected, as discussed in section \ref{sec: assumptions} and 
if the proposed prescription for the natal kicks is correct (which may not be the case if e.g. natal kicks are primarily due to asymmetric neutrino emission),
this case may require a separate treatment.
When we apply the natal kicks following the study of \citet{BrayEldridge16} only after CCSN, we find no significant change with respect to our reference model.
This prescription generally leads to smaller natal kicks during the second supernova and more binaries can survive the explosion, although for
this to have a considerable effect on the number of merging systems one needs more binaries to reach to this stage.
Exactly this is achieved by combining several factors which varied separately do not affect the rates in a noticeable way, but acting together
lead to one of the highest estimates for the merger rates obtained in this study (model $C$).
This shows the importance of investigating also the simultaneous variations of different parameters within the observationally allowed range
while evaluating the evolutionary uncertainties in the merger rates predictions.
\\
\newline
\citet{MennekensVanbeveren14} pointed out that the assumptions regarding the amount of the angular momentum carried away by the material escaping
the binary during the non-conservative mass transfer phases may have a significant effect on the predicted merger rates, finding a difference
 within a factor of $\sim$20 between the DNS rates obtained using their standard prescription with relatively high angular momentum loss and the
 smaller losses closer to the standard assumptions in \textsc{StarTrack}. 
 Here we included a set of models in which we varied the assumptions about the angular momentum loss. 
 The models with extremely high losses (5 times increased with respect to the reference model during the MT with non-compact accretor $J1,3$ and
 twice higher during the MT with NS accretor $J7 \ B$) lead to the highest decrease in the rates that we find in this study and are inconsistent
 with the observational limits.
 The analogous change in the opposite direction - decreasing the angular momentum losses leads to much smaller changes within a factor $\lesssim 4$
 with respect to the reference model.
 The similar estimates obtained for models $J1$ and $J3$ shows that the extent to which the mass transfer is non-conservative
 has a secondary effect on the predicted merger rates when the high angular momentum loss is involved.
 The comparison between cases with conservative, half-conservative and non conservative case with our standard assumptions about the angular momentum loss
 was discussed in \citet{Dominik12}.
\\
\newline
Generally, the models which led to the biggest changes (decrease) with respect to the reference model required quite extreme modifications of the 
standard assumptions (very high angular momentum loss or high natal kicks after the ECS/AIC) in the light of the current knowledge.
This shows that the estimates of the NS-NS merger rates, apart from being particularly sensitive to the treatment of the CE phase, are robust.
\\
\newline
The estimated Galactic merger rates for NS-NS binaries reported by different groups are comparable,
ranging from a few to a few tens Myr$-1$ for the most standard evolutionary assumptions.
In the parameter study performed by \citet{VossTauris03} these rates range from 0.54 - 17 Myr$^{-1}$, with 1.5 Myr$^{-1}$ for their reference model. 
 \citet{MennekensVanbeveren14} report the Galactic rates within the range 0 - 153 Myr$^{-1}$,
which narrows to $\sim$2.5 - 8.5 Myr$^{-1}$ when the standard evolutionary assumptions are used (e.g. their models 2, 22, 23).
In both cases the biggest variations (both the lowest and the highest values) are due to modifications applied to the treatment of the 
common envelope.
\citet{Dominik12} find the range 0.3 - 77.4 Myr$^{-1}$ and 7.6 (23.5) Myr$^{-1}$ for their reference model in variation $B$ ($A$), with the lowest and the highest
values in their study corresponding to the model with conservative mass transfer.
With the NS-NS Galactic merger rates of 24.1 (72.5) Myr$^{-1}$ in variation $B$ ($A$), our reference model falls well within this scheme.
Within our parameter study the rates range from 0.1 - 234 Myr$^{-1}$, where the lowest value corresponds to the
model with extremely high angular momentum loss during MT and the highest was found in our conservative model $J5\ A$  
(see Table \ref{tab:models} for details).\\
It is interesting to note that the detection rates predicted by these groups, as well as our estimate, all
remain at the level of $\sim$1-3 detections per year for the advanced LIGO when the most standard models are considered.
However, different methods were used for their estimation 
\footnote{\citet{VossTauris03} and \citet{MennekensVanbeveren14} applied a simple
scaling rule to calculate the local NS-NS merger rate density based on their Galactic merger rates and assuming a certain density of Milky Way-like
galaxies in the local Universe and that our Galaxy is representative. This local merger rate density was then used to estimate the detection rates.
The detection rates from the study by Dominik et al. were calculated as detailed in \citet{Dominik15}.
They apply a similar procedure to calculate the local merger rate density as the one used in this study 
(we follow the evolution of the NS-NS progenitor systems that form according to the assumed star formation history in chemically
 evolving Universe and merge after a certain delay time, see \ref{ap: rates}) 
and include gravitational waveforms and detector sensitivity curves to convert these cosmological merger rates into detection
rates. They also show the detection rates obtained with the use of eq. 5, which was applied to calculate NS-NS detection rates in this work.}
.
The most optimistic advanced LIGO detection rates predicted within our parameter study yield $\sim$8 - 12 events per year 
of observations (models $NK2$, $C$, $C+P$, see tables \ref{tab:models} and \ref{tab:results}), assuming DNS horizon distance of $\sim$487 Mpc will be reached.
\newline \\
Our study reveals that models that lead to the highest local NS-NS merger rate densities, and thus remain relatively close (within a factor of 2-3) 
to the current LIGO/VIRGO observational limits are often marginally consistent with the constraints on the Galactic NS-NS merger rates
(e.g. models $C$, $C+P$, $NK2$). On the other hand, reference-like models that are in perfect agreement with the estimate given by \citet{Kim15}
produce not enough mergers in the local Universe to meet the constraints imposed by gravitational waves observations.
However, there is no simple translation between the Galactic rate and the cosmological merger rate density.
Our Galactic estimates rely on a single (solar) metallicity, while many different metallicities contribute to the cosmological estimates
and many aspects of stellar evolution are metallicity-dependent (e.g stellar winds). Thus, models leading to similar results in terms
of the Galactic rates may differ significantly when cosmological predictions are compared (and vice versa).
Moreover, constraints based on the observed sample from our Galaxy and those that base on observations of more distant sources that evolved in 
different environments do not necessarily need to agree with each other and thus may not be reconciled with the same model.
\\ \newline
We also note that the merger rate densities estimated for BH-BH binaries in those models that
 meet the constraints imposed by gravitational waves observations span from $\sim$310 - 1000 $\gpy$
 at z$\sim$0 (see table \ref{tab:BHBH rates})
  and reach up to a few $10^{3} \ \gpy$ within the current LIGO range.
 Thus, they exceed the limits of 12-213 $\gpy$ based on gravitational waves observations for this
 type of binaries \citep{Abbott17_GW170104}.
 This might suggest that either different common envelope treatment for DNS and BH-BH progenitors
or/and presence of high BH natal kicks 
\citep[both factors largely affecting the BH-BH merger rates;][]{Belczynski16N} may be needed
to simultaneously meet constraints imposed on both types of binaries.
In particular, models that allow for early binary interaction (with the donor on Hertzsprung gap) 
to be either successful common envelope or thermal timescale mass transfer can 
produce DNS merger rates marginally consistent with LIGO/Virgo estimate,
 while models that have this early interaction lead to a failed common envelope 
 and low BH natal kicks generate the correct BH-BH merger rates. 

\section{Conclusions}

We evaluated the impact of various evolutionary assumptions on the estimated merger rates of the double neutron stars.
Employing the population synthesis method we calculated a suite of physical models
motivated by the most recent theoretical and observational results.
We estimated the Galactic merger rates for a Milky-Way-like galaxy and confronted them with the observational limits based on the properties of the Galactic DNS.
For 17 out of 21 presented models we also show the local merger rate density calculated with a SFR and metallicity varying across the cosmic time as described
in \ref{ap: rates}. These are compared with the limits imposed by gravitational waves and short gamma-ray bursts observations.
Our main conclusions are the following:
\begin{enumerate}
  \item Our reference model predicts Galactic NS-NS merger rates that are fully consistent with the observational limits based on the observed sample of double neutron stars.
	The local merger rate density for this model does not meet the observational constraints
	imposed by the first detection of gravitational waves from coalescing double neutron stars,
	predicting too little NS-NS mergers in the local Universe.
	It also falls slightly below the limits based on short gamma ray bursts 
	that account for their beamed emission. 
	However, as a fraction of short GRBs may originate from neutron star-black hole mergers, our reference model does not contradict these observations.
	
  \item We find only three models that, within the uncertainty limits, 
  are consistent with the current observational constraints from the gravitational waves searches 
  ($C \ A$, $C+P$, $NK2 \ A$, see figure \ref{fig: obs_local_rates}). 
   These models require either applying simultaneous changes in the treatment of common envelope, 
   angular momentum loss, natal kicks and electron-capture supernovae or significantly increasing
    a fraction of core-collapse supernovae that allow for the formation of neutron stars 
    with small ($\lesssim$ 50 km/s) natal kicks.
    At the same time the local BH-BH merger rate densities calculated for those models
    exceed the corresponding observational limits (see table \ref{tab:BHBH rates}).
    This might suggest that different treatment of the common envelope phase for
    BH-BH and NS-NS progenitors or high BH natal kicks are needed to obtain merger
    rates consistent with the observational limits for both types of binaries.
		
  \item We only find two variations that allow to significantly (a factor of $\gtrsim$10) 
  	increase the rates of NS-NS mergers with respect to the reference model.
  	These are the models that combine several factors supporting the formation of
  	double neutron star binaries.
	For the Galactic merger rates the biggest increase is within a factor of $\sim$8 
	(in both cases these are the model $C \ A$ and its other version $C+P$;
	 see tables \ref{tab:models} and \ref{tab:results} for the details).
  
  \item We find several models leading to a significant (more than an order of magnitude) decrease in the predicted DNS coalescence rates.
	These are the models with extremely high angular momentum losses during the stable mass transfer and
	also the models requiring high (in the light of theoretical predictions) magnitudes of natal kick velocities gained
	by neutron stars forming via electron-capture supernovae and/or accretion induced collapse.
	These models are in tension with the observational limits on cosmic rate density of short GRBs (figure \ref{fig: obs_local_rates}) and,
	in case of the models with high angular momentum losses with the limits on the Galactic NS-NS merger rates (figure \ref{fig: obs_galactic_rates}).
  
  \item In many cases the merger rates within the same model can differ by more than an order of magnitude
	depending on the treatment of the common envelope phase with the Hertzsprung gap donor star 
	(see figure \ref{fig: obs_galactic_rates} and the tables).
	This is the most apparent for the models with varied assumptions about the angular momentum loss.
	Such differences are comparable to the largest deviation in the estimated merger rates found between different models studied in this work,
	making the treatment of the common envelope the primary source of evolution-related uncertainties of the merger rates.
	It reveals the need for further investigation of the Roche lobe overflow events with stars that have radiative envelopes and no
	clear density gradient separating the core and the envelope (as in the case of massive Hertzsprung gap stars), in particular the stability conditions.
\end{enumerate}

\section*{Acknowledgements}
We thank to Tomasz Bulik for helpful comments and discussions.
We also thank to Wojciech Gladysz for his help with carrying out the simulations.
The authors acknowledge the Texas Advanced Computing Center (TACC) 
at The University of Texas at Austin for providing HPC
resources that have contributed to the research results reported within this paper (URL: http://www.tacc.utexas.edu).
We also thank to Hannah Brinkman for useful discussion on electron-capture supernovae.
We thank to thousands of {\tt Universe@home} users that have
provided their personal computers and phones for our simulations, and in 
particular to Krzysztof Piszczek and Grzegorz Wiktorowicz (program IT and 
science managers).
KB acknowledges support from the Polish National Science Center
(NCN) grant Sonata Bis 2 (DEC-2012/07/E/ST9/01360).
KB thank the Niels Bohr Institute for its hospitality while part of this
work was completed, and acknowledge the Kavli Foundation and the DNRF for
supporting the 2017 Kavli Summer Program.
MJB acknowledges the support of the NSF through award number HRD-1242090.

%%%%%%%%%%%%%%%%%%%%%%%%%%%%%%%%%%%%%%%%%%%%%%%%%%


\bibliographystyle{mnras}
\bibliography{bibliography} % if your bibtex file is called example.bib

%%%%%%%%%%%%%%%%%%%%%%%%%%%%%%%%%%%%%%%%%%%%%%%%%%

%%%%%%%%%%%%%%%%% APPENDICES %%%%%%%%%%%%%%%%%%%%%

\appendix 
\newpage
 \begin{table*}
\centering
\small
\caption{Short description of the models}
\begin{tabular}{c c c c }
\hline
 \multicolumn{3}{c}{} \\
Model & natal kick & MT & other \\
\multicolumn{3}{c}{} \\
\hline
\hline
\multirow{3}{*}{reference}& std. (CCSN: Hobbs et al. 2005, & std. ($\rm \beta$=1,$\rm f_{a}$=0.5)  &  \\
			  & with $\rm \sigma$=265 km/s  &  &  \\
			  & ECS,AIC: 0)  &  & \\
\hline
\multirow{3}{*}{BE1}& CCSN: Bray \& Eldridge 2016 & std. &  \\
			  & $\rm v_{nk}$=(70 $\rm \frac{M_{ej}}{M_{rem}}$ + 120) km/s  &  &  \\
			  &  ECS,AIC: 0 &  & \\
\hline
\multirow{3}{*}{BE2} & CCSN, ECS: Bray \& Eldridge 2016 & std. &  \\
	    & $\rm v_{nk}$=(70 $\rm \frac{M_{ej}}{M_{rem}}$+ 120) km/s  &  &  \\
			  & AIC: 0  &  & \\
\hline
\multirow{3}{*}{BE3} & CCSN, ECS, AIC: & std. &  \\
		  &  Bray \& Eldridge 2016 &  &  \\
			  & $\rm v_{nk}$=(70 $\rm \frac{M_{ej}}{M_{rem}}$+ 120) km/s  &  & \\
\hline
\multirow{3}{*}{BE4}& CCSN, ECS, AIC: & std. &  \\
			  & Bray \& Eldridge 2016  &  &  \\
			  & $\rm v_{nk}$=(60 $\rm \frac{M_{ej}}{M_{rem}}$+ 70) km/s  &  & \\
\hline
\multirow{3}{*}{BE5}& CCSN: Bray \& Eldridge 2016 & std. &  \\
			  & $\rm v_{nk}$=(60 $\rm \frac{M_{ej}}{M_{rem}}$+ 70) km/s &  &  \\
			  & ECS, AIC: 0  &  & \\
\hline
\multirow{3}{*}{NK1}& CCSN: std. & std. &  \\
			  & ECS: Hobbs et al. 2005, $\rm \sigma$=26.5 km/s &  &  \\
			  & AIC: 0  &  & \\			  
\hline
\multirow{3}{*}{NK2}& CCSN:
$ \left\{ \begin{array}{lr}
    \rm flat & : v_{nk} \leqslant 50 \rm km/s\\
    \rm Hobbs \ et \ al.\ 2005 & : v_{nk} > 50 \rm km/s
  \end{array}\right. $ & &  \\
			  &   & std &  \\
			  & ECS,AIC: 0  &  & \\
\hline
\multirow{3}{*}{EC}	&  	& std. & $\rm M_{cbur1}$=1.63 $\Msun$ \\
			  & std. &  & $\rm M_{cbur2}$=2.45 $\Msun$ \\
			  &   &  & (0.4 $\Msun$ wider range)\\
\hline
\multirow{3}{*}{J1}&  & high J loss $\rm \beta$=5 &  \\
			  & std.  & $\rm f_{a}$=0.5 &  \\
			  &   &  & \\
\hline
\multirow{3}{*}{J2}&  &low J loss $\rm \beta$=0.2 &  \\
			  & std.  & $\rm f_{a}$=0.5 &  \\
			  &  &  & \\			  
\hline
\multirow{3}{*}{J3}&  & high J loss $\rm \beta$=5 &  \\
			  & std. & $\rm f_{a}$=0 &  \\
			  &    & non-conservative & \\
\hline
\multirow{3}{*}{J4}&  &low J loss $\rm \beta$=0.2 &  \\
			  & std.  & $\rm f_{a}$=0 &  \\
			  &  & non-conservative & \\
\hline
\multirow{3}{*}{J5}&  & &  \\
			  & std.  & $\rm f_{a}$=1 &  \\
			  &   & conservative & \\
\hline
\multirow{3}{*}{J6}&  & ang. momentum loss &  \\
			  & std. & during MT with NS &\\
			  &  & 5-times lower &\\	
\hline
\multirow{3}{*}{J7}& & ang. momentum loss &  \\
			  & std. & during MT with NS &\\
			  &   & 2-times higher &\\				  
\hline

\end{tabular}
\label{tab:models}
\end{table*}		  
\makeatletter
\setlength{\@fptop}{0pt}
\makeatother			  
\begin{table*}
\centering
\small
%\caption{Short description of the models}
\begin{tabular}{c c c c }
\hline
 \multicolumn{3}{c}{} \\
Model & kick & MT & other \\
\multicolumn{3}{c}{} \\		  
\hline
\hline
\multirow{3}{*}{W1}&  & &  wind mass-loss rates for\\
			  & std. & std. & H rich stars 50\% lower\\
			  &   &  & for He rich stars 50\% higher \\
\hline
\multirow{3}{*}{W2}&  & &  wind mass-loss rates for\\
			  & std. & std. & H and He rich stars \\
			  &   &  & 50\% higher \\			  
\hline
\multirow{3}{*}{P}&  & & thermal MT instead of CE  \\
			  & std.  & std. &with HG donor  \\
			  &  & &and NS/BH accretor  \\			   	
\hline			   
\multirow{3}{*}{C}& CCSN: Bray \& Eldridge 2016 & low J loss & $\rm M_{cbur2}$=2.45 $\Msun$  \\
			  & $\rm v_{nk}$=(70 $\rm \frac{M_{ej}}{M_{rem}}$+ 120) km/s  &  $\rm \beta$=0.2  & $\rm M_{cbur1}$=1.63 $\Msun$\\
			  & ECS,AIC: 0  & $\rm f_{a}$=0.5 &  \\	
\hline
\multirow{3}{*}{C+P}& CCSN: Bray \& Eldridge 2016 & low J loss & $\rm M_{cbur2}$=2.45 $\Msun$  \\
			  & $\rm v_{nk}$=(70 $\rm \frac{M_{ej}}{M_{rem}}$+ 120) km/s  &  $\rm \beta$=0.2  & $\rm M_{cbur1}$=1.63 $\Msun$\\
			  & ECS,AIC: 0  & $\rm f_{a}$=0.5 & thermal MT instead of CE with HG donor and NS/BH accretor \\	
\hline
\end{tabular}
%\label{tab:models}
\end{table*}
%_________________________
\begin{table*}
\centering
\caption{
Merger rate - measures for different models and metallicities. The first column specifies the model, the second one metallicity.
The third column gives the number of merging DNS systems for solar and two sub-solar metallicities.
The fourth presents the Galactic rates $R_{MW_{Eq}}$ calculated for a Milky Way-equivalent galaxy,
also for the three metallicities. The rates given in brackets are calculated with allowing for CE initiated by HG donors,
while the other rates were calculated with exclusion of these systems (assuming that they do not contribute to the merging population).
For those models where the simulations were performed for all 32 metallicities, the local merger rate density $R_{local}$ is shown in the last column.
In the same column, below the local merger rate density we give the approximate LIGO O3 detection rates $R_{det;\rm \ O3}$ 
calculated using \ref{eq: detection},
assuming the DNS horizon distance of $\sim$385 Mpc (DNS range 170 Mpc) as discussed in section \ref{sec: rates} and
the advanced LIGO detection rates $R_{det; \rm \ adv}$ assuming DNS horizon distance of $\sim$487 Mpc (DNS range 215 Mpc).
The number of simulated ZAMS binaries was $N_{sim}$=2 $\cdot 10^{7}$, unless the other number is provided in the first column. 
\newline
$^{*}$ - local merger rate density was calculated with 2 $\cdot 10^{6}$ ZAMS binaries simulated per metallicity)
}
\begin{tabular}{c c c c c}
\hline
 \multicolumn{4}{c}{}& $R_{local}$ [$\rm Gpc^{-3} \ yr^{-1}$] \\
 Model & Metallicity & N($T_{mr}< $13.7 Gyr) & $R_{MW_{Eq}}$ [Myr$^{-1}$] &   $R_{det;\rm \ O3}$ [$\rm yr^{-1}$] \\
\multicolumn{4}{c}{}& $R_{det; \rm \ adv}$ [$\rm yr^{-1}$]\\
\hline
\hline
\multirow{3}{*}{reference}&  $\rm Z_{\odot}:$ & 8305 [24088]  & 24.1 [72.5] & \multirow{1}{*}{48.4 [190.4]} \\
			   & 10\%$\rm Z_{\odot}:$ &2318 [7693] & 6.5 [22.8] & 0.5 [1.8] \\
			   & 1\%$\rm Z_{\odot}:$& 2694 [15830] & 7.6 [48.3] & 0.9 [3.6]\\
\hline
\multirow{3}{*}{BE1}&  $\rm Z_{\odot}:$ & 10798 [28960] & 30.5 [85.5] & \multirow{1}{*}{87.0 [244.8]} \\
			   & 10\%$\rm Z_{\odot}:$ & 4157 [10317] & 11.5 [30.4] & 0.8 [2.4]\\
			   & 1\%$\rm Z_{\odot}:$& 3736 [17302]& 10.3 [52.3]& 1.7 [4.8]\\
\hline
\multirow{3}{*}{BE2}&  $\rm Z_{\odot}:$ & 434 [4054] & 1.2 [12.3]& \multirow{1}{*}{2.3 [26.0]} \\
		   & 10\%$\rm Z_{\odot}:$ & 218 [805] & 0.7 [2.4]&  0.03 [0.3] \\
			   & 1\%$\rm Z_{\odot}:$& 191 [687]& 0.5 [2.1]& 0.05 [0.6]\\
\hline
\multirow{3}{*}{BE3}&  $\rm Z_{\odot}:$           & 414 [3691]   & 1.3 [11.3]  & \multirow{1}{*}{2.3 [25.9]} \\
			   & 10\%$\rm Z_{\odot}:$ & 215 [836]    & 0.6 [2.5]  & 0.03 [0.3] \\
			   & 1\%$\rm Z_{\odot}:$  & 192 [694]    & 0.6 [2.1]  & 0.05 [0.6] \\
\hline
\multirow{3}{*}{BE4}&  $\rm Z_{\odot}:$           & 885 [8835]   & 2.2 [26.9] & \multirow{3}{*}{-}  \\
			   & 10\%$\rm Z_{\odot}:$ & 400 [1251]   & 1.0 [3.5]  & \\
			   & 1\%$\rm Z_{\odot}:$  & 727 [3010]   & 2.0 [9.1]  & \\
\hline
\multirow{3}{*}{BE5}&  $\rm Z_{\odot}:$           & 8614 [30087]  & 23.4 [88.5]   & \multirow{1}{*}{79.7 [236.7]}  \\
			   & 10\%$\rm Z_{\odot}:$ & 3173 [8737]   & 8.5 [25.1]& 0.8 [2.3]\\
			   & 1\%$\rm Z_{\odot}:$  & 2534 [17195]  & 6.5 [51.9]& 1.6 [4.7] \\
\hline
\multirow{2}{*}{NK1}&  $\rm Z_{\odot}:$ & 359 [1648] & 10.6 [50.8]& \multirow{3}{*}{-}  \\
	    & 10\%$\rm Z_{\odot}:$ &  106 [303] & 3.1 [9.1]& \\
$N_{sim}$=2 $\cdot 10^{6}$ & 1\%$\rm Z_{\odot}:$& 193 [668]& 5.7 [20.6] & \\  
\hline
\multirow{3}{*}{NK2}&  $\rm Z_{\odot}:$ & 9391 [75536] & 26.5 [229.9]& \multirow{1}{*}{45.7 [379.0]}  \\
		    & 10\%$\rm Z_{\odot}:$ &2129 [11678] & 5.9 [35.3]& 0.5 [3.9]\\
		    & 1\%$\rm Z_{\odot}:$& 3013 [35873]& 8.4 [109.6] & 1.0 [7.9]\\				   
\hline
\multirow{3}{*}{EC}&  $\rm Z_{\odot}:$            & 11522 [32589] & 34.0 [98.3] & \multirow{1}{*}{64.5 [250.1]} \\
			   & 10\%$\rm Z_{\odot}:$ & 2888 [10214]  & 8.0 [30.5]  &  0.6 [2.4] \\
			   & 1\%$\rm Z_{\odot}:$  & 4083 [25523]  & 11.7 [78.4]  & 1.3 [4.9]\\
\hline
\multirow{3}{*}{J1}&  $\rm Z_{\odot}:$ & 24 [557]& 0.1 [1.7]& \multirow{1}{*}{1.6 [12.4]} \\
			   & 10\%$\rm Z_{\odot}:$ & 181 [3104] & 0.5 [9.5]& 0.02 [0.1]\\
			   & 1\%$\rm Z_{\odot}:$& 64 [612]     & 0.3 [1.9] & 0.04 [0.3] \\			   			   
\hline
\multirow{3}{*}{J2}&  $\rm Z_{\odot}:$ & 10067 [45532] & 29.3 [137.5]& \multirow{1}{*}{61.7 [256.5]} \\
			   & 10\%$\rm Z_{\odot}:$ & 2664 [8751]& 7.4 [25.7]& 0.6 [2.5] \\
			   & 1\%$\rm Z_{\odot}:$& 3928 [20424]& 11.3 [61.8]& 1.2 [5.0]\\		   			   
\hline
\multirow{3}{*}{J3}&  $\rm Z_{\odot}:$ & 24 [558]    &0.1 [1.6]& \multirow{1}{*}{1.5 [12.0]} \\
			   & 10\%$\rm Z_{\odot}:$ & 181 [3106]& 0.5 [9.5] & 0.02 [0.1]\\
			   & 1\%$\rm Z_{\odot}:$& 64 [614]& 0.1 [1.9]   & 0.03 [0.3]\\
\hline
\multirow{3}{*}{J4}&  $\rm Z_{\odot}:$   &  5980 [24572]& 16.9 [73.7]   & \multirow{3}{*}{-}  \\
			   & 10\%$\rm Z_{\odot}:$ &  1396 [2509] & 4.0 [7.3]   & \\
			   & 1\%$\rm Z_{\odot}:$  & 8547 [9304]  & 26.5 [28.5] & \\			   
 \hline	   
\multirow{3}{*}{J5}&  $\rm Z_{\odot}:$ & 350 [77222] & 1.0 [234.7]& \multirow{1}{*}{57.5 [294.9]} \\
			   & 10\%$\rm Z_{\odot}:$ & 4784 [14685] & 13.4 [42.4]& 0.6 [3.2] \\
			   & 1\%$\rm Z_{\odot}:$&  1099 [8661]& 3.1 [26.4] & 1.3 [6.5]\\
\hline

\end{tabular}
\label{tab:results}
\end{table*}

\makeatletter
\setlength{\@fptop}{0pt}
\makeatother			  
\begin{table*}
\centering
%\caption{}
\begin{tabular}{c c c c c}
\hline
 \multicolumn{4}{c}{} &$R_{local}$ [$\rm Gpc^{-3} \ yr^{-1}$] \\
Model & Metallicity & N($T_{mr}< $13.7 Gyr) & $R_{MW_{Eq}}$ [Myr$^{-1}$] &$R_{det; \rm \ O3}$ [$\rm yr^{-1}$]  \\
\multicolumn{4}{c}{}& $R_{det; \rm \ adv}$ [$\rm yr^{-1}$] \\
\hline
\hline	
\multirow{3}{*}{J6}&  $\rm Z_{\odot}:$ & 5616 [17258] & 15.5 [49.0]& \multirow{3}{*}{-}  \\
			   & 10\%$\rm Z_{\odot}:$ & 2826 [3531] & 8.1 [9.9]& \\
			   & 1\%$\rm Z_{\odot}:$& 4160 [4176]& 12.0 [12.2] & \\
			   \hline
\multirow{3}{*}{J7$^{*}$}&  $\rm Z_{\odot}:$ & 69 [62536] & 0.2 [191.1]& \multirow{1}{*}{16.7 [6.9]}  \\
			   & 10\%$\rm Z_{\odot}:$ & 821 [40109] & 2.0 [122.3]& 0.2 [0.1]\\
			   & 1\%$\rm Z_{\odot}:$& 125 [46946]& 0.4 [143.4] & 0.3 [0.1]\\
\hline
			   
\multirow{3}{*}{W1$^{*}$}	&  $\rm Z_{\odot}:$ & 15487 [29027] & 46.6 [86.9]& \multirow{1}{*}{52.2 [40.3]} \\
			   & 10\%$\rm Z_{\odot}:$ & 2337 [8651] & 6.6 [25.5]& 0.5 [0.4]\\
			   & 1\%$\rm Z_{\odot}:$& 3117 [16776]& 8.9 [51.3] & 1.0 [0.8]\\
\hline
\multirow{3}{*}{W2}&  $\rm Z_{\odot}:$ & 11812 [26860] & 34.7 [80.1]& \multirow{1}{*}{53.9 [187.5]} \\
			   & 10\%$\rm Z_{\odot}:$ & 5911 [15290] & 5.9 [22.3]& 0.5 [1.8]\\
			   & 1\%$\rm Z_{\odot}:$& 2160 [14580]& 6.2 [44.6] & 1.1 [3.6]\\	   			   
\hline
\multirow{1}{*}{P$^{*}$}&  $\rm Z_{\odot}:$ & 16161 [17036] & 96.3 [101.2]& \multirow{1}{*}{89.6 [136.4]}  \\
$N_{sim}$=1 $\cdot 10^{7}$ & 10\%$\rm Z_{\odot}:$ & 12580 [12580] & 74.0 [74.0]& 0.9 [1.3]\\
			   & 1\%$\rm Z_{\odot}:$& 7578 [7578] & 45.0 [45.1] & 1.8 [2.7]\\			   
\hline			   
\multirow{3}{*}{C$^{*}$}&  $\rm Z_{\odot}:$ & 17026 [65236] & 47.9 [193.0]& \multirow{1}{*}{143.3 [448.4]}  \\
			   & 10\%$\rm Z_{\odot}:$ & 11703 [40554] & 33.1 [119.0]& 1.4 [4.4]\\
			   & 1\%$\rm Z_{\odot}:$& 12721 [50320]& 36.2 [149.5] & 2.8 [8.8]\\
\hline			   
\multirow{3}{*}{C+P$^{*}$}&  $\rm Z_{\odot}:$ &  6057 [6700] & 180.5 [199.3]& \multirow{1}{*}{600.5 [631.4]}  \\
			   & 10\%$\rm Z_{\odot}:$ & 3889 [3890] & 115.2 [116.2]&5.7 [6.0]\\
			   & 1\%$\rm Z_{\odot}:$& 3016 [3016] & 90.0 [90.0] &11.5 [12.1]\\
\hline
\end{tabular}
%\label{tab:results}
\end{table*}

 \begin{table*}
\centering
\small
\caption{
Local (z$\sim$0) BH-BH merger rate densities $R_{local;\rm BHBH}$ estimated
for the models that lead to DNS
merger rate densities marginally consistent with the current observational limits imposed by
the first detection of gravitational waves from the merging DNS binary.
The rates given in brackets correspond to submodel $A$, where the common envelope initiated by
a Hertzsprung-gap donor star is allowed, while the other values were calculated with exclusion of
systems that encounter this type of mass transfer.
}
\begin{tabular}{c c c c }
\hline
 Model: & C & C+P & NK2 \\ \hline
 $R_{local;\rm BHBH}$ [$\gpy$]: & 32 [310] & 695 [695] & 190 [1072]\\
\hline
\end{tabular}
\label{tab:BHBH rates}
\end{table*}
%_________________________

\section{Ultra stripped supernovae}\label{ap:USSN}

 Motivated by observations of supernovae with very low peak luminosities ($ \lesssim 10^{42}\;{\rm erg}\,{\rm s}^{-1}$) and rapidly decaying 
 light curves, with a famous example of SN~2005ek \citep{Drout13}, which also holds the record for the smallest ratio of ejecta to remnant mass ever observed for a CCSN,
 \citet{Tauris13} proposed that these events may be due to explosions of ultra-stripped massive stars.
 Ultra-stripped supernovae (USSN) are the explosions of stars which lost nearly entire envelopes during binary interactions with a compact companion.
 The mass of the ejected material during the USSN explosion can be as little as $\lesssim 0.1~\Msun$, they would be thus 'an order of magnitude more stripped'
 than typical progenitors of Ib/Ic SN. The lack of hydrogen in these latter types of SN is usually explained to be due to binary interaction, where
 the initially more massive star was stripped of its outmost layers by mass-transfer and the leftover envelope ejected during the explosion is 
 $\gtrsim 1 \Msun$. 
 Some type~Ib/Ic SNe show much faster than average evolution of the lightcurve, which is interpreted to be due to the fact that their ejecta mass is much smaller,
 they are thus natural candidates for USSN. For further characteristic of the light-curve and spectral properties of ultra-stripped supernovae see \citep{Moriya17}.
\\
 The extreme stripping of the progenitor of the USSN that would leave only the bare stellar core requires a very tight orbit, hence the need for a compact companion.
 It can be achieved in a close, post common envelope X-ray binary, where the second mass-transfer from a naked helium star to a NS,
 initiated by the helium star expansion during its giant phase can occur \citep{Tauris13}.
 It has been postulated by \citet{TaurisLangerPodsiadlowski15} that most, if not all merging double neutron stars form in this scenario,
 with the second NS originating in an ultra-stripped core collapse SN (either iron-core collapse or ECS).
 As argumented by these authors, this formation scenario may potentially lead to small natal kick at the birth of a NS, especially if the mechanism 
 responsible for neutron star natal kicks is primarily due to interaction with asymmetrically ejected material \citep{Janka17}.
 
 \section{Calculation of the merger rates}\label{ap: rates}

In this work we use three different estimates of the merger rates.
The first is the number of double neutron stars merging within the Hubble time for a given model and 
for a given metallicity.
Together with the total stellar mass of the simulations $\rm M_{sim}$ for each model it allows for the
calculation and proper normalization of the rates.
\\ \newline
$\rm M_{sim}$ is the total mass of stars from the whole adopted mass range 0.08-150 $\Msun$
that would be created in order to obtain the $simulated$ number of ZAMS binaries which were chosen from a narrower mass range,
assuming a particular IMF for the mass of the primary star,
primordial mass ratio distribution and the binary fraction $f_{\rm bin}=\frac{N_{\rm bin}}{N_{\rm bin}+N_{\rm single}}$.
This number can be then compared with the stellar mass of the Galaxy, as it is done during calculation of the Galactic merger rates.
We assume that all stars are formed in binaries, which probably overestimates the number of low-mass binaries ($\lesssim 10 \Msun$), but is relevant for massive stars,
as revealed by observational studies 
\citep[$f_{\rm bin}$ = 1, e.g.][]{Sana2012}
Note that this is different than in \citet{Dominik12,Dominik13,Dominik15,Chruslinska17} where $f_{\rm bin}$ = 0.5 was used in the whole mass range
and leads to nearly twice higher overall estimate of the merger rates.
Differences in $\rm M_{sim}$ for different models come from differences in the number of simulated ZAMS binaries for these models.

\subsection{Galactic merger rates}

Secondly, we present the Galactic merger rates i.e. the number of coalescences per unit time within a galaxy.
These are calculated for a fiducial Milky Way-like galaxy, assuming 10 Gyr of continuous star formation at the rate of SFR$_{\rm gal}$=3.5 $\Msun$/yr.
The Galactic rates are calculated for a single metallicity.
The procedure is summarized below.\\
Each of the successfully formed DNS binaries is assigned a random birth time between 0 and 10 Gyr (the Galactic age $T_{\rm gal}$).
If its delay time, being the sum of the time from the birth of the binary to the formation of the second compact object ($\sim$Myrs)
and the time required for its coalescence (due to orbital evolution caused by gravitational waves emission, $\gtrsim$ 100 Myr), is close
to the current age of the galaxy, its added to the number of coalescing systems $N_{\rm mr}$.
One has to properly normalize the obtained number to match the total mass of the Galaxy formed in stars during its lifetime ($T_{\rm gal}\cdot$SFR$_{\rm gal}$),
as in general the mass within the simulated binaries corresponds to only a fraction of the needed mass.
This is done by simply using each of the simulated binaries $n$=$T_{\rm gal}\cdot$SFR$_{\rm gal}$/$\rm M_{sim}$ times for the calculation of $N_{\rm mr}$.

\subsection{The local merger rate density}

As a third estimate for those models where the simulations were run for a wide spectrum of metallicities we
provide the local merger rate density (number of coalescences per unit time per unit volume in the local Universe, i.e. around redshift=0),
calculated with taking into account the star formation rate and metallicity evolution with redshift.
We use the method described in \citet{BelczynskiRepetto16} (their \begin{it} method III \end{it}) but present the 'total rates', without the division for the
redshifted total mass bins.
Furthermore, we adopt different cosmic star formation rate SFR(z) as in \citet{Belczynski16N}, following the study of \citet{Madau14} 
based on far-UV and IR observations:
\begin{equation}\label{eq:SFR(z)}
 \rm SFR(z) = 0.015 \frac{(1+z)^{2.7}}{1+[(1+z)/2.9]^{5.6}} \ \Msun Mpc^{-3} yr^{-1}
\end{equation}
and metallicity evolution with redshift as provided by these authors,
but increased by 0.5 dex to better fit observational data \citep{Vangioni15}:
\begin{equation}\label{eq:Z(z)}
%\begin{split}
 \rm log(Z_{mean}(z)) = 0.5 + \\ \rm log\left( \frac{y(1-R)}{\rho_{b}} \int_{z}^{20} \frac{97.8 \times 10^{10} SFR(z')}{H_{0}\sqrt{\Omega_{M}(1+z')^{3}+\Omega_{k}(1+z')^{2}+\Omega_{\Lambda}}} dz'\right)
%\end{split}
 \end{equation}
where $Z_{mean}$ is the mean metallicity at the given redshift, R=0.27 is the return fraction (fraction of stellar mass returned into the interstellar medium),
y=0.019 is the metal yield (mass of metals liberated into the interstellar medium by stars per unit mass locked in stars), SFR(z) is given by eq. \ref{eq:SFR(z)}
and the cosmological parameters are: $\rho_{b}=2.77 \times 10^{11} \ \Omega_{b}h_{0}^{2} \
\Msun \rm Mpc^{-3}$ (baryon density), $\Omega_{b}=0.045$, $h_{0}=0.7$, $\Omega_{\Lambda}=0.7$,$\Omega_{M}=0.3$, $\Omega_{k}=0$, $H_{0}=70 \rm \ km \ s^{-1} Mpc^{-1}$.
We model binaries for 32 metallicities ranging from 0.5\% $\Zsun$ to 1.5$\cdot \Zsun$ (see sec. Methods in \citet{Belczynski16N} for the full list).
The merger rate density as a function of redshift is calculated by integrating through the cosmic star-formation (SFR) as described in \citet{BelczynskiRepetto16}
using redshift bins $\Delta$z = 0.1 up to z = 15.\\
It is worth to point out some implicit assumptions about the IMF related to this procedure.
We assume that the IMF does not depend on metallicity, using the same function regardless of the metallicity or redshift.
This is justified by the fact that studies based on the integrated properties of the observed galaxies and resolved stellar populations so far revealed
no clear evidence for any significant variations or systematic trends in the form of the IMF with redshift or metallicity \citep{Massey03,Bastian10,Offner14}, 
apart from local deviations \citep[e.g. top-heavy IMF found in some globular clusters, see][]{Kroupa13}.
Secondly, to obtain the SFR in the form given by eq. \ref{eq:SFR(z)}, \citet{Madau14} assumed a Salpeter-like IMF \citep{Salpeter55}
with the constant slope of $\alpha_{\rm IMF}=2.35$ for masses ranging from 0.1 - 100 $\Msun$.
Hence, the use of eq. \ref{eq:SFR(z)} together with our results relying on a Kroupa-like IMF is inconsistent.
Klencki et al. (in prep.) with the use of the \textsc{Starburst99} code \citep{Leitherer99,Leitherer14} calculated the correction factor $\rm K_{\rm IMF}$=0.51 
which can be used as a multiplicative factor in equation \ref{eq:SFR(z)} to calculate the rate density in a more consistent manner
(see section 2 and appendix therein).
We thus modify the SFR according to their results, using: $\rm SFR_{\rm Kroupa}(z) = K_{\rm IMF} \times SFR(z)$.

\subsubsection{Uncertainties involved in the calculations}\label{ap: rate_err}

There is a number of poorly constrained components within our
cosmological calculations that are responsible for the buildup of
the final uncertainty associated with our estimated NS-NS merger rate density.
Here we evaluate the impact of some of them on our predictions, leaving a more detailed discussion for future studies.
\\
The most important source of uncertainty in the estimated NS-NS merger rate density is the form of the IMF used
(different choices of the IMF are present in the literature, e.g. \citealt{Salpeter55, Kroupa1993, Kroupa01,Chabrier03})
and in particular the slope of the high-mass end of the initial mass function for field stars,
where observational datapoints are scarce \citet[e.g. 2 in][]{Bastian10}.
Using a steeper $\alpha_{IMF}$=2.7 for primary stars more massive than the Sun,
as used e.g. by \citep{Dominik12} and applying the appropriate SFR correction (Klencki et al. in prep) 
would increase the estimates by a factor of $\sim$1.25, 
while less steep high-mass end of the IMF would decrease the NS-NS local merger rate density
 (e.g. $\alpha_{IMF}$=1.9 would reduce the estimate by a factor of $\sim$0.6).
Any further break in the form of the IMF beyond 10 $\Msun$, which presently cannot be excluded
 would affect our predictions for NS-NS (with steeper IMF further increasing the rates).
\\
The assumed IMF should be used consistently with the adopted star formation rate density,
which is not well determined at higher redshifts z$\gtrsim$2 (the SFR prescription used in this study may
underestimate the number of stars formed at high redshifts; see \citealt{Belczynski16PISN},
 although this is of secondary importance for the calculation of the local NS-NS merger rates
considered in this work) and with the metal yields $y$ and return fractions $R$.
As discussed by \citep{Vincenzo16}, the latter two depend not only on the
particular choice of the IMF, but also on the upper cut-off mass $m_{up}$ of the IMF and on the
input set of the stellar yields used in the calculations (comparing two sets they suggest
a factor of 1.5 uncertainty related to this choice).
However, the impact of the uncertainties that affect both $y$ and $R$ on our results partially cancels out,
as our calculations are sensitive only to the term $y(1-R)$ present in eq. \ref{eq:Z(z)}.
The more top-heavy the IMF, the more the predicted yields vary with
the location of the upper mass cut-off (increase with increasing mass).
We put this upper limit on $m_{up}$ =150 $\Msun$ and use yields calculated for
$m_{up}$ =100 $\Msun$, potentially underestimating this value.
The use of twice higher average metal yields would result in a factor of $\sim$1.2 increase
in the predicted NS-NS merger rate density.
\\
Our result is also sensitive to the assumed binary fraction.
Here we assumed that all stars form in binaries, adopting $f_{bin}$=1 in the whole simulated mass range
and possibly overestimating the number of low-mass binaries.
Different choice of $f_{bin}$=0.5 would result in a factor of $\sim$1.5 increase in NS-NS merger rate density.
Most likely different binary fraction should be used for high and low-mass primaries, however it is not
clear where lies the boundary between the low and high mass regime in this matter,
furthermore this transition may turn out to be more gradual.
Note that this choice would affect e.g. the predicted ratio of NS-NS and BH-BH mergers.

\section{Typical evolution of the progenitor of the merging DNS}\label{ap: evolution}

\begin{figure}
	\includegraphics[scale=0.3]{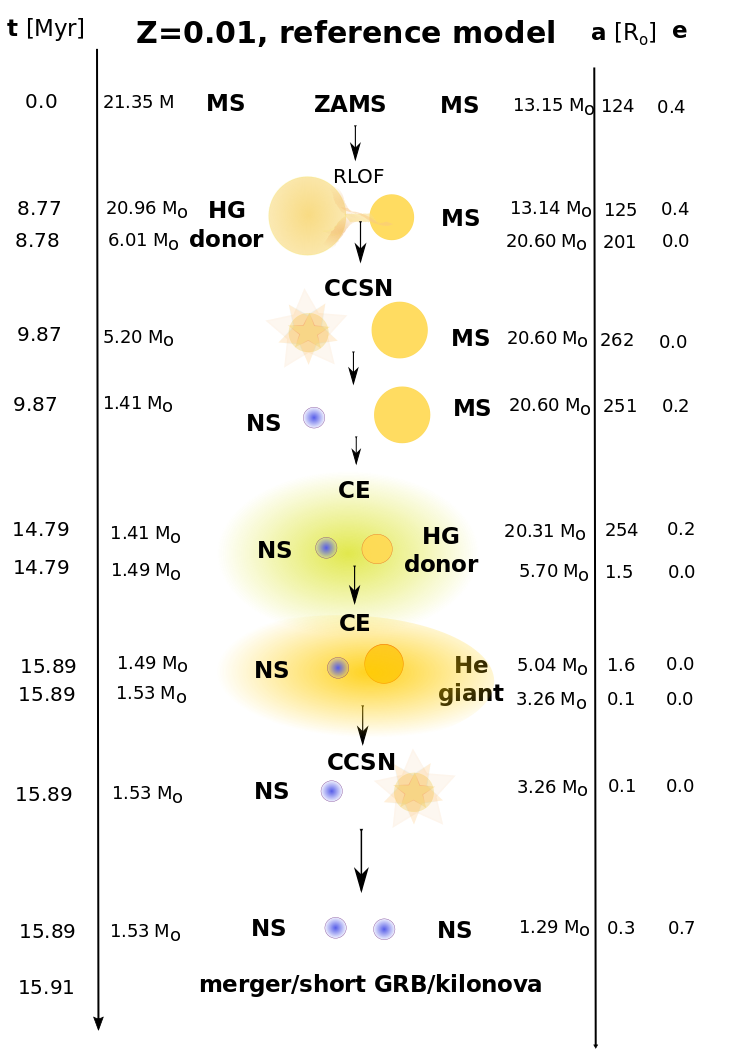}
    \caption{
    Example evolution of a merging double neutron star with the remnant masses
      similar to the masses of GW170817 progenitor stars (see \citealt{1stNSNS_merger_paper}),
      formed in a channel involving two common envelope
      episodes (CE) and shown for the reference model (see sec. \ref{sec: ref_model}) and metallicity $Z$=0.01.
      Stars pass through a stable Roche lobe overflow (RLOF) phase before the first core-collapse event.
      The first star undergoes a core-collapse supernova (CCSN) and forms a neutron star (NS).
      The binary passes through the common envelope phase (CE) initiated by a Hertzsprung-gap donor (HG) star,
      shrinking the orbit and leaving behind a helium star, which later on again overfills its Roche lobe
      and initiates another phase of dynamically unstable mass transfer. 
      The second stars collapses to NS and a double neutron star system is formed, which then
      gradually decreases its separation due to emission of gravitational waves until both stars merge,
      leading to a strong gravitational wave signal, short gamma-ray burst (GRB) and kilonova emission. 
      Note that the whole evolution of the binary in this case lasts only $\sim$16 Myrs.
      This evolutionary channel (involving two CE phases) produces binaries that merge shortly after the formation of the second NS.
      }
    \label{fig:GW170817}
\end{figure}

As can be seen in the figure \ref{fig:ref02}, only a small fraction of the simulated ZAMS binaries (drawn from the limited mass range)
ends up forming a double neutron star system that will merge within the Hubble time.
Nearly half of the initial binaries undergoes an early CE phase ($\sim$1/3 is HG-donor case).
Majority of these systems is lost in mergers before the first compact object can form.
The biggest contribution to the merging population from this 'early-CE' channel comes from binaries passing through the HG-donor CE 
and for the reference model ($\Zsun$) is at the level of $\sim 10^{-3}$\% of the simulated ZAMS binaries.
Around 20 \% of the simulated systems does not interact before the formation of the first remnant. Their contribution to the merging population of DNS
is negligible, as they form binaries that are too wide to merge within the Hubble time.\\
The remaining 1/3 of the binaries interact via only stable MT before the formation of either WD, BH or NS (with the majority forming a WD,
as a consequence of the IMF favoring the formation of the less massive stars).
The significant fraction of the potential NS progenitors is lost at the moment of the first supernova explosion.
The combination of the abrupt mass loss and the natal kick velocity causes $\sim$ 80\% of the binaries undergoing the SN to disrupt at this point.
Over 99 \% (in submodel $B$; 95 \% in submodel $A$) of the surviving binaries that will end up as merging DNS form through ECS 
(with no natal kick added in the reference model).\\
In case of the binaries that remain bound after the SN and where the first component finishes its evolution as a NS, one can distinguish
between the systems that will encounter no CE and will not contribute to the merging population and those that will undergo the common envelope evolution.
A fraction of the potential DNS progenitors coalesce during the CE phase, before the envelope can be ejected.
1/3 of those that survive passes through the highly uncertain CE initiated by the Hertzsprung-gap donor.
In the analyzed model their contribution to the merging population of DNS is the biggest, being twice the number of the interesting
binaries resulting from the evolution through the other types of CE.
In $>87 \%$ (depending on the model) of merging DNS progenitors the late highly non-conservative mass transfer (due to exceeded Eddington rate) after the CE initiated by an expanding naked helium star occurs,
potentially leading to further stripping of the envelope and subsequent SN explosion with very little ejected material. 
The discussed evolutionary path is summarized with an example in the figure \ref{fig: detailed}.
For sub-solar metallicities the fraction of binaries that encounter this late mass transfer decreases 
and another variation of the discussed channel, requiring two unstable mass transfer episodes,
becomes dominant. Figure \ref{fig:GW170817} shows an example of this variation.
Again a fraction of binaries does not form a DNS, as the secondary star ends up as a WD or BH, even more get disrupted during the supernova explosion.
A comparable number of the merging DNS forms in the scenario requiring an accretion-induced collapse of the star that initially formed a WD to a NS
and the later common envelope phase, although we find that the great majority of these (merging) systems passes through the CE initiated by the HG-donor,
once more pointing out the need for better understanding the physics of mass transfer phases with this kind of donor stars.
\\
The NS-NS binary shown in example in figure \ref{fig:GW170817} was chosen to reproduce the known properties
of GW170817 (see the limits reported by \citealt{1stNSNS_merger_paper}).
Note that most of merging DNS forming in our simulations require the first NS to form with a small natal kick and thus they preferentially form
through electron-capture SN. 
However, ECS produce rather lightweight NS ($\sim1.3 \Msun$) and to reach a relatively high mass of the primary star as in GW170817 ($\sim1.5 \Msun$),
NS would have to accrete a few $\sim$0.1 $\Msun$.
At the same time neutron stars are believed to accrete little mass during the subsequent mass transfer episodes \citep[$<$0.1 $\Msun$, e.g. sec.4 in][]{Tauris17_DNS},
which makes the ECS formation scenario unlikely. 
As NS accrete primarily during common envelope phase, the formation channel with two CE favors the formation of more massive neutron stars.
It should be noted that the amount of matter accreted during the common envelope is not well constrained.
We assume the accretion rate of 10\% of the Bondi-Hoyle rate. However recent calculations show that the accretion rates can be reduced even by a factor of $10^{-2}$
with respect to the rates resulting from the standard Bondi-Hoyle approximation when the structure of the envelope is taken into account \citep{McLeod15a,MacLeod17,Murguia-Berthier17}.

%%%%%%%%%%%%%%%%%%%%%%%%%%%%%%%%%%%%%%%%%%%%%%%%%%

% Don't change these lines
\bsp	% typesetting comment
\label{lastpage}
\end{document}